\documentclass[journal]{IEEEtran}

\ifCLASSINFOpdf
\else
\fi

\usepackage[dvipdfm]{graphicx}
\usepackage{graphicx}
\usepackage{amssymb}
\usepackage{amsmath}
\usepackage{amsfonts}
\usepackage{scalefnt}
\usepackage{color}
\usepackage{cite}
\usepackage{multicol}
\usepackage{theorem}
\usepackage{hyperref}
\usepackage{amsmath}
\usepackage{comment}
\usepackage{float}
\usepackage[caption=false]{subfig}
\usepackage{algorithm}
\usepackage{algorithmic}
\usepackage[dvipsnames]{xcolor}
\begin{document}
\bstctlcite{IEEEexample:BSTcontrol}
\title{Novel KLD-based Resource Allocation for Integrated Sensing and Communication}
\author{Yousef Kloob,~\IEEEmembership{Member,
~IEEE}, Mohammad Al-Jarrah,~\IEEEmembership{Member,
~IEEE}, Emad Alsusa,~\IEEEmembership{Senior Member,~IEEE,} and Christos Masouros,~\IEEEmembership{Senior Member,~IEEE}\vspace{-0.4in}\thanks{%
Y. Kloob, M. Al-Jarrah, E. Alsusa are with the Department of Electrical and
Electronic Engineering, University of Manchester, Manchester M13 9PL, U.K.
(e-mail: \{yousef.kloob, mohammad.al-jarrah, e.alsusa\}@manchester.ac.uk), and C. Masouros is with Department of Electronic and Electrical Engineering, University College London, London WC1E 6BT, U.K (e-mail: c.masouros@ucl.ac.uk).}\thanks{%
}}


\maketitle

\begin{abstract}
In this paper, we introduce a novel resource allocation approach for integrated sensing-communication (ISAC) using the Kullback–Leibler divergence (KLD) metric. Specifically, we consider a base-station with limited power and antenna resources serving a number of communication users and detecting multiple targets simultaneously. First, we analyze the KLD for two possible antenna deployments, which are the separated and shared deployments, then use the results to optimize the resources of the base-station through minimising the average KLD for the network while satisfying a minimum predefined KLD requirement for each user equipment (UE) and target. To this end, the optimisation is formulated and presented as a mixed integer non-linear programming (MINLP) problem and then solved using two approaches. In the first approach, we employ a genetic algorithm, which offers remarkable performance but demands substantial computational resources; and in the second approach, we propose a rounding-based interior-point method (RIPM) that provides a more computationally-efficient alternative solution at a negligible performance loss. The results demonstrate that the KLD metric can be an effective means for optimising ISAC networks, and that both optimisation solutions presented offer superior performance compared to uniform power and antenna allocation.

\end{abstract}


\IEEEpeerreviewmaketitle

\section{Introduction}
\IEEEPARstart{T}{he} field of communication technology is undergoing rapid evolution, with 6G being the next generation of wireless communication networks that promises to bring about transformative changes \cite{9145564,9040264,9144301}. As we approach this new era of wireless communication, the demand for advanced and efficient communication techniques has become increasingly apparent \cite{rajatheva2020white}. 6G networks are expected to support various services and applications, such as holographic communications, smart grids, digital twins, and integrating radar technology into communication systems which is the interest of this paper. 

Integrated sensing and communication (ISAC), when properly implemented, is exceptionally beneficial, as both systems could use the same hardware and network resources. For instance, the large antenna arrays which form massive multiple-input-multiple-output (mMIMO) can be shared among both subsystems, which reduces the cost immensely \cite{7898445,8288677,8386661,9705498,9226446,9385108,9540344}. Generally speaking, ISAC systems can be implemented in two ways, namely, the separated deployment in which the base station (BS) antennas are distributed among each sub-system and the shared deployment in which antennas are exploited for both sub-systems. Additionally, incorporating both systems and utilising the same spectrum bands leads to improved system capacity, higher data rates and better sensing capabilities due to increased spectral efficiency \cite{article}. Furthermore, radar systems are generally used for detection, localisation, tracking, and navigation, which have the potential to provide precise data about the surrounding environment, and thus numerous applications such as autonomous vehicles and drones can benefit from such information to take appropriate actions in order to accomplish their tasks \cite{8554265,8753528,8706972}.

One of the challenging factors which should be carefully considered for effective ISAC functionality is the allocation of available resources at the BS. Remarkable efforts have been devoted in the literature to achieve this objective. For instance, the optimisation of power allocation while maximising the signal-to-interference-noise-ratio (SINR) for the radar system or the communication system capacity, either as cooperative or coexistence design, has been discussed in literature \cite{9359480,8834831,9764299,9593118}. However, the measures considered in these articles to evaluate the performance of the communication and sensing subsystems are not homogeneous, i.e., different measures for each subsystem. Consequently, it is difficult to assess the performance gained by the power optimisation on the functionality of the overall system.

In \cite{9359480,8834831}, power allocation and transmission strategies are proposed with the aim of optimising the performance of coexisting radar and communication systems. The work in \cite{9359480} focuses on orthogonal frequency division multiplexing (OFDM)  communication systems and sparse sensing radars in cluttered environments and develops a joint transmit design framework accordingly, where the SINR of the radar is maximised with constraints on the throughput of the communication system. The optimisation problem relies on prior knowledge of the communication system signal-to-noise ratio (SNR), interference-to-noise-ratio (INR), and radar system clutter-to-noise-ratio (CNR), and the target SNR up to an unknown constant common to all subcarriers. In \cite{8834831}, the authors present an optimisation framework that maximises the performance metric of one system while meeting the service constraint of the other system where both radar-centric and communication-centric designs are studied. OFDM dual functional radar and communication (DFRC) based systems that serve the dual purpose of radar and communication are considered in \cite{9764299} and \cite{9593118}. The authors in \cite{9764299} have presented an optimal power allocation method for monostatic OFDM systems by maximising the mutual information (MI) between the radar target and communication users while minimising the radar detection error probability. Moreover, \cite{9593118} proposes an algorithm that balances the competing requirements of radar detection and communication data rate, aiming at optimising the power and maximising performance of the DFRC system.

Apparently, none of these papers explored the joint allocation antenna and power resources in both of the antenna deployment configurations for ISAC systems and the comparison between the two models (e.g. shared and separated deployments), and how these models affect the transmission and reception of both the communications and radar. However, incorporating the optimal antenna allocation for each subsystem into the problem adds a new dimension to the problem where power and antennas for each system are controllable to achieve the highest performance. In addition, ISAC systems aim at integrating the communication and sensing tasks in a more holistic manner than DFRC or a coexistent operation.
\begin{figure*}[ht]
\vspace{-0.2in}
\centering
\includegraphics[width=6.4in]{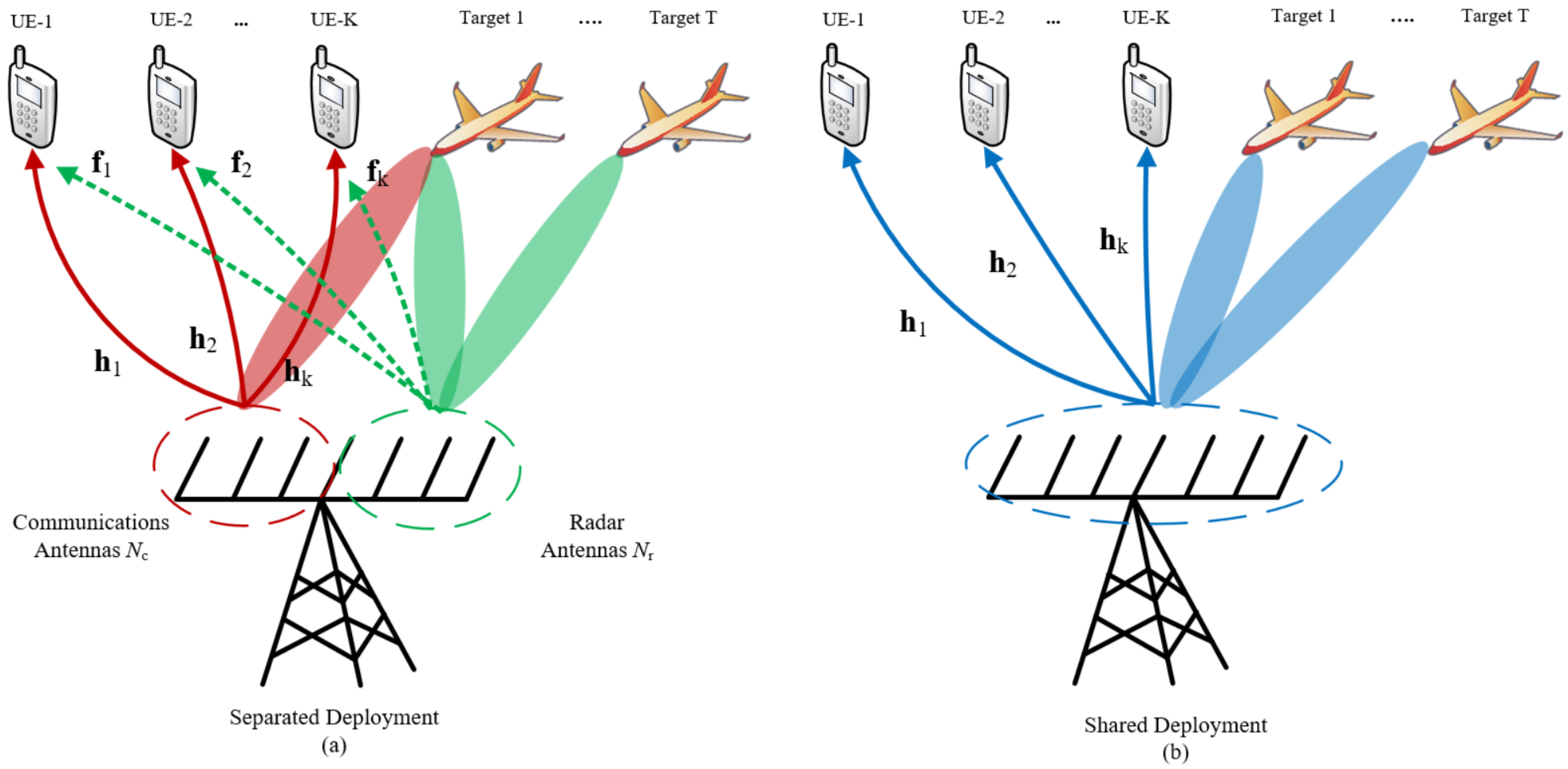}
\vspace{-0.1in}
\caption{Antenna deployments for ISAC systems. (a) Separated deployment; (b) Shared deployment.}
\label{fig:combined}
\vspace{-0.1in}
\end{figure*}
Interestingly, in our previous works \cite{Al-Jarrah2023,10118838}, we have shown that a unified performance evaluation measure using Kullback–Leibler divergence (KLD) can be effectively used to characterise ISAC systems holistically. The relation between KLD from one side and the symbol error rate (SER) and detection performance of the communication system and radar system, respectively, from the other side, has been thoroughly investigated. Moreover, the trade-off between both subsystems using the introduced KLD has been well studied.  Obviously, KLD can be utilised as a design reference for the overall ISAC system as a single entity and put both subsystems on one scale, unlike conventional designs, which rely on two different measures, i.e., one measure for each subsystem. Furthermore,  unlike SER, detection probability and false alarm probability, KLD measure is independent of the detection process and detection threshold applied at the receiver side, which makes it an ideal candidate for system level design and resource allocation \cite{7086341,4776572}.

Inspired by the benefits of KLD, this paper investigates utilizing such metric for optimal resource allocation in ISAC networks. We do not only focus on a typical ISAC setup, but also explore the models used in \cite{8288677,9200993}, for shared and separated deployment modes. As shown in Fig. 1.a, in shared deployment, communication and radar systems use the entire antenna array, and therefore optimization is only for power allocation. In contrast, the separated deployment in Fig. 1.b, in which each subsystem is allocated a certain subset of antennas, requires the optimization of both power and antenna allocation for communication and radar systems.
We present an optimisation framework for power and antenna allocation for each node, i.e., communication user or sensing target, which results in the maximum achievable network-level KLD while satisfying a minimum KLD requirement that needs to be satisfied for each device. The formulated problem is found to be a constrained non-convex mixed integer non-linear programming (MINLP) for the separated deployment, and nonlinear programming (NLP) for the shared deployment. Therefore, we propose a low-complexity algorithm for solving the separated deployment optimisation problem based on the interior-point method (IPM), which is referred to rounding-based interior-point method (RIPM) algorithm. The achievable performance of the proposed algorithm is compared to a genetic algorithm (GA) as a benchmark, as well as, both algorithms are compared with the uniform resource allocation (URA) scenario as an additional benchmark. The results show the effectiveness of the proposed algorithm when compared to GA and URA, where the obtained results demonstrate that both GA and RIPM significantly outperform the URA scenario. Moreover, as compared to GA, the proposed RIPM provides comparable detection capability with more computational efficiency.
The primary contributions of this paper are summarized as follows:
\begin{comment}

\begin{itemize}
    \item We formulated a system model that seamlessly integrates clutter for both separated and shared deployments in ISAC systems.
    
    \item We derived the KLD specifically for radar and communication systems within the shared deployment ISAC system.
    
    \item We proposed a new approach to power and antenna allocation using KLD, framing the problem as a constrained non-convex MINLP for separated deployment and as an NLP for shared deployment.
    
    \item We developed the RIPM algorithm based on the interior-point method (IPM) to \textcolor{ForestGreen}{address} the MINLP optimisation challenge.
    
    \item We conducted a detailed comparative analysis between RIPM, GA, and URA, emphasizing computational efficiency and convergence rates.
    
    \item We provided empirical validation that underscores the performance benefits of RIPM and GA over the URA methodology.
    
    \item We evaluated the efficacy of both shared and separated deployment methods, drawing attention to the superior performance outcomes of shared deployment when assessed using the KLD metric.
\end{itemize}
\color{red}
\end{comment}
\begin{itemize}
    \item A generalized system model with multiple communication users and multiple targets is introduced, which seamlessly integrates the impact of clutter and imperfect interference cancellation. In addition, both the separated and shared ISAC antenna deployments are investigated thoroughly.
    
    \item Providing the analysis of the KLD to characterize the achievable detection capability for radar and communication systems within the shared deployment of ISAC systems.
    
    \item Proposing a novel approach for power and antenna allocation using the derived KLD. The formulated optimization problems are found to be  constrained non-convex MINLP for the separated deployment and as a constrainted non-convex NLP for the shared deployment. A low complex optimization algorithm, that includes RIPM algorithm based on the IPM to solve the MINLP, is developed. While the shared deployment NLP is solved using IPM. 
    
    \item A detailed comparative analysis between RIPM, GA, and URA, are conducted to emphasize the computational efficiency and convergence rates.
    
    \item Extensive simulated validations are provided that underscore the performance benefits of RIPM and GA over the URA methodology. Furthermore, the efficacy of both shared and separated deployment methods is evaluated, drawing attention to the superior performance outcomes of shared deployment when assessed using the KLD metric.
\end{itemize}
\color{black}
In summation, this research underscores the relevance of KLD as an effective metric for system optimization in ISAC systems. The results shows that RIPM algorithm matches the performance of the established GA, with significant reduction in the computational complexity, where also, both the GA and RIPM methods surpass the URA in effectiveness. Importantly, when comparing the shared and separated deployment methods, the shared deployment yielded superior results based on the KLD measure. This shows the potential benefits of progressing towards a more unified ISAC system.

The remainder of this paper is organised as follows. In section II, the system model is presented. Section III introduces a summary of the derivation of the radar KLD, the communication system's KLD, and the average KLD for the overall ISAC system. In section IV, the optimisation problem for power and antenna allocation is formulated. Section V shows the numerical results. Section VI the complexity analysis for the optimisation techniques are explored and Section VII concludes the work.
\\
\textit{$Notation$}:  The following notations are used in this paper. Bold uppercase letters (e.g. $\mathbf{S}$) specify matrices, while
bold lowercase letters (e.g. $\mathbf{s}$) specify vectors. Superscripts ${(\cdot)}^*$, ${(\cdot)}^T$ and ${(\cdot)}^H$ specify the conjugate, transpose and Hermitian
transpose, respectively. Subscripts ${(\cdot)}_\mathrm{c}$ and ${(\cdot)}_\mathrm{r}$ relate the
corresponding parameter to the communication or radar system, respectively. The absolute value and the trace operator are denoted by $|\cdot|$ and $ \mathrm{tr}\{\cdot\}$, respectively.

\section{System Model}
This paper is concerned in a multi-user multi-target (MUMT) MIMO scenario. The ISAC system model comprises an $N$ antenna MIMO-BS, these antennas are utilized for detecting ${T}$ number of radar targets, and serving $K$ number of single-antenna communication user equipments (UEs) in downlink. The ISAC system is analyzed for two system deployments, the separated and shared deployment as shown in Fig.\ref{fig:combined}.

The total transmitted power available at BS is $P_\mathrm{T}$ which is utilised for both sensing and data communication functionalities. The portions of the power that are allocated to the radar and communication subsystems are respectively denoted as $P_\mathrm{r}$ and $P_\mathrm{c}$, where $P_\mathrm{T}=P_\mathrm{c}+P_\mathrm{r}$. Zero-forcing (ZF) beamforming is employed at BS to precode the information of communication UEs and eliminate the cross-interference caused by signals intended for other UEs \cite{Fatema2018MassiveML}. Moreover, the radar waveform is designed such that the covariance matrix is $\mathbf{R}_{x}\triangleq \frac{1}{L}\sum_{l=1}^{L}\mathbf{x}_{l}\mathbf{x}_{l}^{H}\in \mathbb{C}^{N_\mathrm{r}\times N_\mathrm{r}}$, where $L$ is the total number of snapshots, and $\mathbf{x}_l\in \mathbb{C}^{N_\mathrm{r} \times 1}$ is the radar waveform vector for the $l$-th snapshot. 
\vspace{-0.05in}
\subsection{Separated Deployment}

In this case, as shown in Fig.\ref{fig:combined}.a for the separated deployment antenna model, the MIMO-BS antennas are divided into two subsets. The first subset contains a number of $N_\mathrm{r}$ antennas applied for detecting ${T}$ number of radar targets, whereas the second one utilises the remaining $N_\mathrm{c}=N-N_\mathrm{r}$ to serve $K$ number of single-antenna communication user equipments (UEs) in the downlink. 
\subsubsection{Communication System}

At each $l$ instance, a data symbol $s_{k,l}$ intended for the $k$-th UE is drawn from a certain normalised constellation, i.e, $\mathbb{E}[ \left\vert s_{k,l} \right\vert ^{2}] =1$. Given the channel matrix from MIMO-BS to UEs, $\mathbf{H}_l$, these symbols, i.e., $s_{k,l}\forall k$, are precoded using a ZF precoder with a precoding matrix $\mathbf{W}_{\mathrm{c},l}\in \mathbb{C}^{N_{\mathrm{c}}\times K}$ that is normalised using instantaneous matrix normalisation scheme for which $\mathbf{W}_{\mathrm{c},l}=\frac{\mathbf{\Tilde{W}}_{\mathrm{c},l}}{\sqrt{\mathbf{s}_{l}^{H}\:\mathbf{\Tilde{W}}_{\mathrm{c},l}\:\mathbf{\Tilde{W}}_{\mathrm{c},l}^{H}\:\mathbf{s}_{l}}}$ with $\mathbf{\Tilde{W}}_{\mathrm{c},l}= \mathbf{H}_{l}^{H}\:\left(\mathbf{H}_{l}\:\mathbf{H}_{l}^{H}\right)^{-1}$ is the non-normalised ZF precoding matrix. This sort of normalisation ensures that the communication system transmit power satisfies the power constraint. The received signal $\mathbf{y}_{l} \in \mathbb{C}^{K \times 1}$, at the $l$-th instance, can be shown as follows,
\begin{equation}
\mathbf{y}_{l} =\mathbf{D}_\mathrm{c}\:\mathbf{H}_{l}^{T}\:\mathbf{W}_{\mathrm{c},l}\:\mathbf{P}_\mathrm{c}\:\mathbf{s}_{l}\mathbf{+}\:\mathbf{D}_\mathrm{c}\:\sqrt{\frac{P_{\mathrm{r}}}{N_{\mathrm{r}}}}\mathbf{F}_l^{T}\mathbf{x}%
_{l}\mathbf{+}\mathbf{n}_l, \label{eq:1}
\end{equation}
where $\mathbf{P}_\mathrm{c}=\mathrm{diag}(P_{\mathrm{c},1},P_{\mathrm{c},2},\cdots,P_{\mathrm{c},K})$ is a power control matrix for UEs, $\mathbf{H}_{l}\in \mathbb{C}^{N_{\mathrm{c}}\times K}\sim \mathcal{C}\mathcal{N}%
\left( 0,2\sigma _{H}^{2}\right) $ is the BS-UEs communication channel matrix which is modeled as flat Rayleigh fading, $\mathbf{F}_l^{T} \in \mathbb{C}^{N_{\mathrm{r}}\times T}\sim \mathcal{C}%
\mathcal{N}\left( 0,2\sigma _{F}^{2}\right) $ is the radar-communication interference channel matrix modeled as flat Rayleigh fading from the radar antennas to each of the UEs, and $\mathbf{n}_l\in \mathbb{C}^{K\times 1}\sim \mathcal{C}\mathcal{N}\left( 0,2\sigma _{n}^{2}\right) $ is the additive white Gaussian noise (AWGN). The matrix $\mathbf{D}_\mathrm{c}=\mathrm{diag}(d_{\mathrm{c},1}^{-\eta/2},d_{\mathrm{c},2}^{-\eta/2},\cdots,d_{\mathrm{c},K}^{-\eta/2})$ counts for the free space pathloss, where $\eta$ is the pathloss exponent, and $d_{\mathrm{c},k}$ is the $k$-th UE distance to BS. Throughout the paper, the channels are assumed to be independent and identically distributed (iid).
\subsubsection{Radar System}

For this system, we are targeting the scenario in which each target is located in a separate radar bin \cite{https://doi.org/10.1049/joe.2019.0456,9229166}, allowing for easy identification for the number of targets in the environment. We further assume that BS has knowledge about the number of possible targets in the area through historical records. MIMO radar makes it possible to generate multiple beams simultaneously by combining multiple orthogonal signals. Accordingly, the transmitted signal vector at the output of the antennas can be represented as,
\begin{equation}
\mathbf{x}_{l}=\mathbf{W}_{\mathrm{r},l}\:\mathbf{P}_\mathrm{r}\:\Phi, \label{eq:2}
\end{equation}\color{black}
where $\Phi =\left[ \phi _{1},\phi _{2},\cdots ,\phi _{T}\right] ^{T}$ is a set of $T$ orthonormal baseband waveforms \cite{5419124}, $\mathbf{P}_\mathrm{r}=\mathrm{diag}\left(\sqrt{P_{\mathrm{r},1}/{N_\mathrm{r}}},\sqrt{P_{\mathrm{r},2}/{N_\mathrm{r}}},\cdots,\sqrt{P_{\mathrm{r},T}/{N_\mathrm{r}}}\right)$ is the power allocation matrix which is used to control the amount of power to be emitted towards each target, and $\mathbf{W}_{\mathrm{r},l}\in \mathbb{C}^{N_{\mathrm{r}}\times T}$\ is the precoding matrix for the radar at the $l$-th signalling period. The precoding matrix for the radar system can be tailored to enhance the radar performance or fulfil a specific radar covariance matrix requirement. For instance, a radar covariance matrix with desired characteristics can be achieved through the design of appropriate precoding matrix ${\mathbf{R}}_{\mathrm{w}} \triangleq \frac{1}{L}\mathbf{W}_{\mathrm{r}}\times \mathbf{W}_{\mathrm{r}}^H=\mathbf{I}_{N_\mathrm{r}\times N_\mathrm{r}}$ which is typically used for omnidirectional radar. 

The radar return from all targets is processed through a bank of matched filters with the signal waveform $\phi _{t}  \: \forall t=\left\{ 1,2,\cdots ,T\right\} $ which is tuned to a certain radar angular-range-Doppler bin. Thereafter, since $\phi_{t} \perp \phi_i  \: \forall t \neq i$, the radar return from different targets can be separated, and thus the detection of each target can be performed independently. The binary hypothesis problem associated with each target can be defined as $\mathcal{H}_{q} \: \forall q\in\{0,1\}$, where $q=0$ and $q=1$ denote the absence and presence of a target, respectively. Mathematically, the received radar signal from target $t$ under hypothesis $\mathcal{H}_{q}$ can be written as
\begin{equation}
\mathbf{y}_{\mathrm{r},t,l|\mathcal{H}_{q}} =\sqrt{\frac{
P_{\mathrm{r},t}}{N_{\mathrm{r}}}}d_{\mathrm{r},t}^{-\eta/2}\:\alpha_t\:\textbf{A}\left( \theta_\mathrm{t} \right) 
\mathbf{w}_{\mathrm{r},t,l}\:q+\mathbf{\tilde{\omega}}_{\mathrm{r}} ,
\label{eq:3}
\end{equation}%
where $d_{\mathrm{r},t}^{-\eta}$ is the two-way channel pathloss from BS to the target and backwards with $d_{\mathrm{r},t}$ is the two-way distance, $\alpha_t$ represents the target cross-section, and $\textbf{A}\left( \theta_t \right) =\textbf{a}_{\mathrm{T}}\left( \theta_{t}\right)\times\textbf{a}_{\mathrm{R}
}\left( \theta _{t}\right) $ is the equivalent array manifold with $\textbf{a}_{\mathrm{T}}\left( \theta_{t}\right)$ and $\textbf{a}_{\mathrm{R}}\left( \theta_{t}\right)$ represent the transmit and receive steering vector for the $t$-th target, respectively. In this paper, it is assumed that $\textbf{a}\left( \theta_{t}\right) \triangleq \textbf{a}_{\mathrm{T}}\left( \theta_{t}\right)=\textbf{a}_{\mathrm{R}}\left( \theta_{t}\right)$. After employing interference cancellation (IC) at BS, the noise plus interference residue due to the communication signal is denoted as $\mathbf{\tilde{\omega}}_{\mathrm{r}}\triangleq \mathbf{\omega }_{\mathrm{r}}+\mathbf{n}_{\mathrm{r},l} \sim \mathcal{CN}\left( 0,2\sigma _{\tilde{\omega}}^{2}\mathbf{I}_{N_{\mathrm{r}}}\right) $, with total variance $\sigma _{\tilde{\omega}}^{2}=\sigma _{\omega }^{2}+\sigma _{n}^{2}$, where $\mathbf{n}_{\mathrm{r},l}\in \mathbb{C}^{N_\mathrm{r}\times 1}\sim \mathcal{C}\mathcal{N}\left( 0,2\sigma _{n}^{2}\:\mathbf{I}_{N_\mathrm{r}}\right)$ is AWGN with $\mathbf{I}_{N_\mathrm{r}}$ is an identity matrix. Here, $\mathbf{\omega }_{\mathrm{r}}\in \mathbb{C}^{N_{\mathrm{r}}\times1}\triangleq \mathbf{G}_{\mathrm{err}}\:\mathbf{P}_\mathrm{c}\:\mathbf{W}_{\mathrm{c},l}\:\mathbf{s}_{l}$ represents the residual error of the imperfect IC that is employed to mitigate interference caused by the communication signal, with $\mathbf{G}_{\mathrm{err}}\in \mathbb{C}^{N_\mathrm{r}\times N_\mathrm{c}}$ is the channel estimation error matrix at BS. The residual error of imperfect IC, $\mathbf{\omega }_{\mathrm{r}}$, is approximated to a complex Gaussian distribution by assuming that the channel estimation errors are normally distributed with each entry in $\mathbf{G}_{\mathrm{err}}$ being $\mathcal{CN}\left( 0,2\sigma_\mathrm{err }^{2}\right)$, where $2\sigma_\mathrm{err }^{2}$ is the variance of the channel estimator. Each element of $\mathbf{\omega }_{\mathrm{r}}$ is a sum of independent $KN_\mathrm{c}$ random variables. Therefore, central limit theorem (CLT) can be applied to approximate the density of the elements of $\mathbf{\omega }_{\mathrm{r}}$ for large $KN_\mathrm{c}$. This approximation leads to the conclusion that the errors caused by imperfect IC are complex Gaussian distributed. Specifically, $\mathbf{\omega }_{\mathrm{r}}$ follows a complex Gaussian distribution with mean zero and covariance matrix $2\sigma_{\omega}^{2}\mathbf{I}_{N_\mathrm{r}}$, where $\sigma_{\omega }^{2}=\sigma_{\mathrm{err}}^{2}\sigma_{\mathrm{w}}^{2}N_{\mathrm{c}}\:P_\mathrm{c}$. Here, $\sigma_{\mathrm{w}}^2$ is the variance of the elements of $\mathbf{w}_{\mathrm{c},k,l}$, which are the elements of the precoding matrix $\mathbf{W}_{\mathrm{c},l}\triangleq[\mathbf{w}_{\mathrm{c},1,l},...,\mathbf{w}_{\mathrm{c},k,l}]^T$.

\subsection{Shared Deployment}
In this scenario, as shown in Fig.\ref{fig:combined}.b, the MIMO-BS antennas are not divided into two subsets for the shared deployment model. Instead, the whole the antenna array with number of elements $N$ is utilized by both systems for detecting ${T}$ number of radar targets, and serving $K$ number of single-antenna communication user equipments (UEs) in downlink. It is important to notice that the transmitted signal in shared deployment is a combined superimposed signal of both radar and communication signals, which can be represented as
\begin{equation}
\mathbf{\dot{W}}_{\mathrm{c},l}\:\mathbf{P}_\mathrm{c}\:\mathbf{s}_{l}+\mathbf{\dot{x}}_{l}, \label{eq:shh}
\end{equation}
where $\mathbf{\dot{x}}_{l}=\mathbf{\dot{W}}_{\mathrm{r},l}\:\mathbf{P}_\mathrm{r}\:\Phi$ has similar properties to the communication signal of the separated deployment in \eqref{eq:2} but with different dimensions as the shared deployment utilizes $N$ antennas rather than $N_\mathrm{r}$. Therefore, the precoding matrix for the communication and radar systems are $\mathbf{\dot{W}}_{\mathrm{c},l}\in \mathbb{C}^{N\times K}$, respectively, and $\mathbf{\dot{W}}_{\mathrm{r},l}\in \mathbb{C}^{N\times T}$, and the power control diagonal matrix for the radar system is $\mathbf{\dot{P}}_\mathrm{r}=\mathrm{diag}\left(\sqrt{P_{\mathrm{r},1}/{N}},\sqrt{P_{\mathrm{r},2}/{N}},\cdots,\sqrt{P_{\mathrm{r},T}/{N}}\right)$.
\subsubsection{Communication system}
The received signal $\mathbf{y}_{l} \in \mathbb{C}^{K \times 1}$, at the $l$-th instance, can be shown as follows,
\begin{equation}
\mathbf{y}_{l} =\mathbf{D}_\mathrm{c}\:\mathbf{\dot{H}}_{l}^{T}\:\mathbf{\dot{W}}_{\mathrm{c},l}\:\mathbf{P}_\mathrm{c}\:\mathbf{s}_{l}+\mathbf{D}_\mathrm{c}\:\sqrt{\frac{P_{\mathrm{r}}}{N}}\:\left(\mathbf{\dot{H}}_{l}^{T}+\mathbf{\dot{F}}^{T}\right)\:\mathbf{\dot{x}}_{l}%
\mathbf{+}\mathbf{n}_{l}, \label{eq:1sh}
\end{equation}
 the first term in the received signal represents the useful part of the transmitted signal after the going through the fading channel and the pathloss, while the second term represents the radar interference from clutter and the transmit signal. $\mathbf{\dot{H}}_{l}\in \mathbb{C}^{N\times K}$ is the BS-UEs communication channel matrix which is modeled as flat Rayleigh fading,, and $\mathbf{\dot{F}}_l^{T} \in \mathbb{C}^{N\times T}$is the radar-communication interference channel matrix modeled as flat Rayleigh fading from the radar antennas to each of the UEs, and $\mathbf{n}_l\in \mathbb{C}^{K\times 1}\sim \mathcal{C}\mathcal{N}\left( 0,2\sigma _{n}^{2}\right) $ is the additive white Gaussian noise (AWGN). The clutter term, noise and the transmit signal radar interference can be encapsulated as $\mathbf{\zeta} \in \mathbb{C}^{K\times1} \triangleq \mathbf{D}_\mathrm{c}\:\sqrt{\frac{P_{\mathrm{r}}}{N}}\:\left(\mathbf{\dot{H}}_{l}^{T}+\mathbf{\dot{F}}^{T}\right)\:\mathbf{x}_{l}\mathbf{+}\mathbf{n}_{l} $ each element of $\zeta$ has a variance of $\sigma _{\zeta }^{2}=d_{\mathrm{c},k}^{-\eta}P_{\mathrm{r}}\sigma_{\dot{H}}^{2}+d_{\mathrm{c},k}^{-\eta}P_{\mathrm{r}}\sigma_{\dot{F}}^{2}+\sigma _{n}^{2}$.

\subsubsection{Radar system}
The binary hypothesis problem associated with each target can be defined as $\mathcal{H}_{q} \: \forall q\in\{0,1\}$, where $q=0$ and $q=1$ denote the absence and presence of a target, respectively. Mathematically, the received radar signal from target $t$ under hypothesis $\mathcal{H}_{q}$ can be written as
\begin{multline}
\mathbf{y}_{\mathrm{r},t,l|\mathcal{H}_{q}} =
\alpha_t\:\textbf{A}\left( \theta_\mathrm{t} \right)\left(\sqrt{\frac{P_{\mathrm{r},t}}{N_{}}}d_{\mathrm{r},t}^{-\eta/2}\mathbf{\dot{w}}_{\mathrm{r},t,l}\:q+\mathbf{\dot{W}}_{\mathrm{c},l}\:\mathbf{P}_\mathrm{c}\:\mathbf{s}_{l}\right)\\
+\mathbf{G}_\mathrm{err_1}\:\left(\sqrt{\frac{P_{\mathrm{r}}}{N_{}}}\:\mathbf{\dot{x}}_{l}\:q+\mathbf{\dot{W}}_{\mathrm{c},l}\:\mathbf{P}_\mathrm{c}\:\mathbf{s}_{l}\right)+\mathbf{n}_{\mathrm{r},l}.
\label{eq:shrad}
\end{multline}
 The first part of the signal in parentheses is the transmitted signal after reception, while the second term in parentheses is the clutter signal reflected back to the BS. Similar to the derivation of \eqref{eq:3}, \eqref{eq:shrad} can be reduced to 
\begin{equation}
\mathbf{y}_{\mathrm{r},t,l|\mathcal{H}_{q}} =\sqrt{\frac{
P_{\mathrm{r},t}}{N}}d_{\mathrm{r},t}^{-\eta/2}\:\alpha_t\:\textbf{A}\left( \theta_\mathrm{t} \right) 
\mathbf{\dot{w}}_{\mathrm{r},t,l}\:q+\mathbf{\tilde{\zeta}},
\end{equation}
where $\tilde{\zeta}=\alpha_t\:\textbf{A}\left( \theta_\mathrm{t} \right)\:\mathbf{W}_{\mathrm{c},l}\:\mathbf{P}_\mathrm{c}\:\mathbf{s}_{l} + \mathrm{G_{err_1}}\:\left(\sqrt{\frac{P_{\mathrm{r}}}{N_{}}}\mathbf{x}_{l}\:q+\mathbf{W}_{\mathrm{c},l}\:\mathbf{P}_\mathrm{c}\:\mathbf{s}_{l}\right)+\mathbf{n}_{\mathrm{r},l} $ is the interference from the clutter, plus the interference from the combined transmitted signal, and the noise. There is two operations of imperfect IC that are important to note, one that involves the clutter part of the signal which follows the same explanation done before in the separated deployment model and is represented with the channel estimation error matrix at the BS $\mathbf{G}_\mathrm{err_1}\in \mathbb{C}^{N\times N} \sim \mathcal{CN}\left( 0,2\sigma_\mathrm{err_1}^{2}\right) $. While the other IC involves the interference from the transmitted signal itself which is a combined signal of both systems, this IC depends on the estimation of the target cross section $\alpha_t$ and the equivalent array manifold $\textbf{A}\left( \theta_\mathrm{t} \right)$, where the estimation is considered imperfect where we considered a combined Gaussian error $\mathbf{G}_\mathrm{err_2}\in \mathbb{C}^{N\times N} \sim \mathcal{CN}\left( 0,2\sigma_\mathrm{err_2}^{2}\right) $ with variance for both of the parameters represented as $\sigma_\mathrm{err_2}^2$. Following these approximations leads to the conclusion that all the imperfect IC errors along with the complex gaussian noise can be combined into a complex gaussian with zero mean and $2\sigma_{\tilde{\zeta}}^{2}\mathbf{I}_{N}$ covariance matrix, where $\sigma_{\mathbf{\tilde{\zeta}}}^2=
\sigma_{\mathrm{n}}^2+\sigma_{\mathrm{err_1}}^2\:\left(\sigma_{\mathrm{\dot{w}}}^2\:N\:\mathrm{P}_\mathrm{c}+\mathrm{tr}\{R_x\}\:\mathrm{P}_\mathrm{r}\:/N\right)+
\sigma_{\mathrm{err_2}}^2\:\sigma_{\mathrm{\dot{w}}}^2\:N\:\mathrm{P}_\mathrm{c}$. Here, $\sigma_{\mathrm{\dot{w}}}^2$ is the variance of the elements of $\mathbf{\dot{w}}_{\mathrm{c},k,l}$, which are the elements of the precoding matrix $\mathbf{\dot{W}}_{\mathrm{c},l}\triangleq[\mathbf{\dot{w}}_{\mathrm{c},1,l},...,\mathbf{\dot{w}}_{\mathrm{c},k,l}]^T$.
\vspace{-0.1 in}
\section{Kullback–Leibler divergence}

This section presents the derivation of KLD for both the separated and shared deployment which is to be used in the formulation of the optimisation problem for both systems.
\vspace{-0.06 in}
\subsection{Separated Deployment}
\subsubsection{KLD fo Communication System}
In the case of ZF precoding, the precoding matrix is given in the section introducing the communication system. The KLD shown below accommodates any type of constellations and has been modified to account for the long-term pathloss effect as follows,
\vspace{-0.05 in}
\begin{equation}
\mathrm{KLD}_{\mathrm{c},k} =\frac{\lambda \:\alpha _{\mathrm{ZF}}^{2}\:P_{\mathrm{c},k}\:d_{\mathrm{c},k}^{-\eta}}{2\:M\left( M-1\right) \left(P_{\mathrm{r}}\: \sigma _{F}^{2}\:d_{\mathrm{c},k}^{-\eta}+\sigma _{n}^{2}\right)\ln2},
\label{eq:7}
\end{equation}%
where $M$ is the modulation order, $\alpha _{\mathrm{ZF}}=\sqrt{N_\mathrm{c}-K+1}$ is the normalisation factor for ZF precoding, and $\lambda$ is a constant that depends on the constellation, for example, in the case of $M$-ary Phase Shift Keying (M-PSK) it would be $\lambda_{\mathrm{MPSK}}=M^2$. 

\subsubsection{KLD for Radar System}
The sufficient statistics of the generalised likelihood ratio test is denoted as $\xi (\theta_k)$. As $L$ increases, $\xi (\theta_k)$ distribution approaches a Chi-squared distribution \cite[Eq. 54]{7089157}. Hence, $\xi (\theta_\mathrm{t})$ can be expressed as,
\begin{equation}
\mathrm{\xi }\left( \theta _{t}\right) \sim \left\{ 
\begin{array}{l}
\mathcal{H}_{1}:\mathcal{X}_{2}^{2}\left( \lambda _{t}\right)  \\ 
\mathcal{H}_{0}:\mathcal{X}_{2}^{2}\left( 0\right) 
\end{array}%
\right.  , \label{eq:8} 
\end{equation}%
where $\mathcal{X}_{2}^{2}\left( \lambda _{t}\right) $ denotes a noncentral Chi-squared random variable with $2$ degrees of freedom and a noncentrality parameter of
\begin{equation}
\lambda_{t}=\frac{\alpha_t\:d_{\mathrm{r},t}^{-\eta}\:P_{\mathrm{r},t}\:|\textbf{a}^H(\theta_{t})\:\textbf{R}_{t}\:\textbf{a}(\theta_{t})|^2}{N_{\mathrm{r}}\:(\sigma_{n}^2+\sigma_{\mathrm{err}}^{2}\sigma_{\mathrm{w}}^{2}N_{\mathrm{c}}\:P_\mathrm{c}\:d_{\mathrm{r},t}^{-\eta})}, \label{eq:9}
\end{equation}
where $\mathbf{R}_{t}=\frac{1}{L}\sum_{l=1}^{L}\mathbf{w}_{\mathrm{r},t,l} \mathbf{w}_{\mathrm{r},t,l}^{H}$.which can be reduced to $\lambda _{t}=\dfrac{ \alpha_t\:N_{\mathrm{r}}\:d_{\mathrm{r},t}^{-\eta}\:P_{\mathrm{r},t}}{(\sigma_{n}^2+\sigma_{\mathrm{err}}^{2}\sigma_{\mathrm{w}}^{2}N_{\mathrm{c}}\:P_\mathrm{c}\:d_{\mathrm{r},t}^{-\eta})}$ for the case of orthogonal waveforms.

The KLD for the radar from $\mathrm{\xi }_{\mathcal{H}_1}$ to $\mathrm{\xi }_{\mathcal{H}_0}$ is found to be \cite{Al-Jarrah2023},

\begin{equation}
\mathrm{KLD}_{ \mathrm{\xi }_{\mathcal{H}_{0}}\! \parallel  \mathrm{\xi }_{\mathcal{H}_{1}} \!} %%%%%%%%%%%%%%%%%%%%%%%%%%%%%%%%%%%%%%%%%%
\!=\!\!\frac{1}{2}\!\left(\! 1.4427\lambda _{t}\!-\!\frac{1}{\ln 2}%
\!\!\int\limits_{0}^{\infty }\!\!\!\mathrm{e}^{-0.5\xi }\ln \left( I_{0}\left(\!\! \sqrt{%
\lambda _{t}\xi }\!\right) \!\right)\! \right)\!\! d\xi , \label{eq:10} 
\end{equation}
where $I_{0}(.)$ is a modified bessel function of the first kind. That is, the KLD for the radar from $\mathrm{\xi }_{\mathcal{H}_0}$ to $\mathrm{\xi }_{\mathcal{H}_1}$ is derived as,

\begin{equation}
\mathrm{KLD}_{\mathrm{\xi }_{\mathcal{H}_{1}}\parallel \mathrm{\xi }_{\mathcal{H}_{0}}}=\frac{-0.5\lambda _{t}}{\ln 2}+\frac{\mathrm{e}^{-0.5\lambda _{t}}}{2\ln 2}\mathcal{I}, \label{eq:11} 
\end{equation}
where $\mathcal{I}\!=\!\!\int\limits_{0}^{\infty }\mathrm{e}^{-0.5\xi }I_{0}\left( 
\sqrt{\lambda _{t}\xi }\right) \ln \left( I_{0}\left( \sqrt{\lambda _{t}\xi }%
\right) \right) d\xi $ can be solved with numerical methods such as trapezoidal integration. Consequently, for the $t$th target, the average KLD can be written as $\mathrm{KLD}_{\mathrm{r},t}\!\!=\!0.5\left(\:\mathrm{KLD}_{ \mathrm{\xi }_{\mathcal{H}_{0}}\parallel  \mathrm{\xi }_{\mathcal{H}_{1}}}+\mathrm{KLD}_{\mathrm{\xi }_{\mathcal{H}_{1}}\parallel \mathrm{\xi }_{\mathcal{H}_{0}}}\right)$.
\subsection{Shared Deployment}

\subsubsection{KLD for the Communication system}
The KLD here differs from separated deployment as the radar interference is essentially doubled as shown in \eqref{eq:1sh}, as there is two sources of radar interference, the first is the  transmit signal radar interference, and the second is from the radar clutter. For multivariate Gaussian distributed random variables having mean
vectors of $\mathbf{\mu }_{m}$ and $\mathbf{\mu }_{n}$\textbf{\ }and
covariance matrices of\textbf{\ }$\Sigma _{m}$ and $\Sigma _{n}$, the KLD can be
derived as\vspace{-0.075in}
\begin{multline}
\mathrm{KLD}_{n\rightarrow m}=\frac{1}{2\ln 2}\left( \mathrm{tr}\left(
\Sigma _{n}^{-1}\Sigma _{m}\right) -2+\left( \mathbf{\mu }_{k,n}-\mathbf{\mu 
}_{k,m}\right) ^{T} \right. \\
\; \left.  \times\;\Sigma _{n}^{-1} \left( \mathbf{\mu }_{k,n}-\mathbf{\mu }%
_{k,m}\right) +\ln \frac{\left\vert \Sigma _{n}\right\vert }{\left\vert
\Sigma _{m}\right\vert }\right).\label{zb}\vspace{-0.04in}
\end{multline}

By noting that $\Sigma _{n}=\Sigma _{m}=\sigma _{\zeta }^{2}\mathbf{I}_{2}$,
and given that $\mathbf{\mu }_{k,m}=\left[ \sqrt{P_{\mathrm{c},k}}\:d_{\mathrm{c},k}^{-\eta/2}\alpha
_{\mathrm{ZF}}\cos \phi _{k,m},\sqrt{P_{\mathrm{c},k}}\:d_{\mathrm{c},k}^{-\eta/2}\alpha _{\mathrm{ZF}}\sin \phi _{k,m}\right] $, and $\mathbf{\mu }_{k,n}=\left[ \sqrt{P_{\mathrm{c},k}}\:d_{\mathrm{c},k}^{-\eta/2}\alpha
_{\mathrm{ZF}}\cos \phi _{k,n},\sqrt{P_{\mathrm{c},k}}\:d_{\mathrm{c},k}^{-\eta/2}\alpha _{\mathrm{ZF}}\sin \phi _{k,n}\right] $ Then by substituting $\Sigma _{n}$, $\Sigma _{m}$, $\mathbf{\mu }_{k,m}$, and $\mathbf{\mu }_{k,n}$ in \eqref{zb}, then the KLD for shared deployment communication system can be found as\cite[Corrolary 1]{Al-Jarrah2023},

\begin{equation}
\mathrm{KLD}_{\mathrm{c},k} =\frac{\lambda \:\alpha _{\mathrm{ZF}}^{2}\:P_{\mathrm{c},k}\:d_{\mathrm{c},k}^{-\eta}}{2\:M\left( M-1\right) \left(P_{\mathrm{r}}\: \sigma _{H}^{2}\:d_{\mathrm{c},k}^{-\eta}+P_{\mathrm{r}}\: \sigma _{F}^{2}\:d_{\mathrm{c},k}^{-\eta}+\sigma _{n}^{2}\right)\ln2},
\label{eq:7}
\end{equation}%
\subsubsection{KLD for the Radar System}

After collecting $L$ snapshots, the received signal matrix can
be formulated as

\begin{eqnarray}
\!\!\!\mathbf{Y}_{\mathrm{r},t|\mathcal{H}_{q}} \!\!&\!\!=\!\!&\!\!\sqrt{\frac{P_{%
\mathrm{r},t}}{N}}\alpha _{t}\textbf{A}\left( \theta_{t} \right) \!\mathbf{W}_{%
\mathrm{r},t}\:q\!+\!\mathbf{\tilde{Z}}_{\mathrm{r}},
\label{eq:sh_rad}
\end{eqnarray}
where $\mathbf{{W}}_{t,\mathrm{r}}\in\mathbb{C}^{N\times L}=%
\left[ \mathbf{w}_{t,\mathrm{r},1} ,\mathbf{w}_{t,\mathrm{r},2%
} ,\cdots\right.$$\left.,\mathbf{w}_{t,\mathrm{r},l}\right] 
$ and $\mathbf{\tilde{Z}}_{\mathrm{r}}\in \mathbb{C}^{N\times L}=\left[ 
\mathbf{{\tilde{\zeta}}}_{t,\mathrm{r},1} ,\mathbf{%
{\tilde{\zeta}}}_{t,\mathrm{r},2} ,\cdots ,\right.$$\left. \mathbf{%
{\tilde{\zeta}}}_{t,\mathrm{r},l} \right] $. By
noting that $\mathbf{y}_{\mathrm{r},t|H_{1}}\left[ l\right] \sim 
\mathcal{CN}\left(\sqrt{\frac{P_{\mathrm{r}}}{N}}\alpha
_{t}\mathbf{A}\left( \theta _{t}\right) \mathbf{w}_{\mathrm{r},t,l} ,2\sigma _{{\tilde{\zeta}}}^{2}\mathbf{I}_{N}\right) $, then using this the KLD can be derived from \eqref{zb} using \eqref{eq:sh_rad} as follows,

\begin{multline}
    \mathrm{KLD_{r,t}}=\frac{|\alpha _{t}|^2\:P_{\mathrm{r},t}\:d_{\mathrm{r},t}^{-\eta}\:|a(\theta_t)\:\mathrm{R_x}\:a(\theta_t)^H|^2}{2\:N\:\sigma_{\tilde{\zeta}}^2\:\ln2}
\end{multline}

By considering the derived KLD for all radar targets and all served communication UEs, the weighted average KLD for the overall ISAC network can be formulated as

 \begin{equation}
\mathrm{KLD}_{\mathrm{avg}} =\:\left(\mathrm{c_{com}}\sum_{k=1}^{K}{\mathrm{KLD}_{\mathrm{c},k}}+\mathrm{c_{rad}}\sum_{t=1}^{T}{\mathrm{KLD}_{\mathrm{r},t}}\right),
\label{eq:13}
\end{equation}%
where $\mathrm{c_{com}}$ and $\mathrm{c_{rad}}$ are the weights for each system, with $K\:\mathrm{c_{com}}+T\:\mathrm{c_{rad}}=1$. The weights gives priority to the desired system which allows for more flexibility and trade-off.
\section{Power and Antenna Allocation}
Effective power and antenna allocation is a critical design factor for ISAC systems, as it directly impacts the system's sensing and communication capabilities. As well as, allocating more resources to one system does not only reduce the resources of the other subsystem, but also could result in a considerable amount of interference which may lead to system breakdown. Nevertheless, by optimising these parameters through maximizing the system's $\mathrm{KLD}_{\mathrm{avg}}$, a significant improvement on the detection of radar targets and the quality of communication among UEs can be achieved. This, in turn, results in better overall system performance and effective resource utilisation. In this section, we formulate two optimisation problems for separated deployment and shared deployment. It is noteworthy mentioning that the shared deployment utilizes all of the antennas for both systems, therefore, the required computational resources are reduced by eliminating the need for allocating the antennas, while in the separated deployment the optimization algorithm needs to take into account the allocation of both antennas and power.

\subsection{Separated Deployment}

The optimisation problem is formulated to maximise the average KLD by optimally allocating a limited budget of power and antennas among UEs and radar targets while satisfying certain KLD requirements for every served device (e.g. communication UEs and sensed targets). With the aid of \eqref{eq:13}, the optimisation problem can be formulated as follows,
\begin{subequations}
\begin{align}
\mathcal{P}_{1}:\max_{P_{\mathrm{c},k},P_{\mathrm{r},t},N_{\mathrm{c}},N_{\mathrm{r}}} \quad\!\! & \mathrm{KLD}_{\mathrm{avg}}(P_{\mathrm{c},k},P_{\mathrm{r},t},N_{\mathrm{c}},N_{\mathrm{r}}) \label{10.a}\\
\textrm{s.t.}\quad &  A_{t,\mathrm{L}}\leq{\mathrm{{KLD}}_{\mathrm{r},t}}\leq A_{t,\mathrm{U}},  \,\,\,  \forall t \in T,  \label{10.b}\\
  &B_{k,\mathrm{L}}\leq{\mathrm{{KLD}}_{\mathrm{c},k}}\leq B_{k,\mathrm{U}},      \forall k \in K,   \label{10.c}  \\
  & \sum_{k=1}^{K}{P_{\mathrm{c},k}}+\sum_{t=1}^{T}{P_{\mathrm{r},t}}= P_{\mathrm{T}},  \label{10.d}\\
  & {N_{\mathrm{c}}}+{N_{\mathrm{r}}}= N.  \label{10.e}
\end{align}
\end{subequations}
where $ A_{t,\mathrm{L}}$ is the minimum KLD for the $t$-th radar target, and $B_{k,\mathrm{L}}$ is the minimum KLD for the $k$-th UE, an upper limit for the KLD can also be enforced with the use of $ A_{t,\mathrm{U}}$, and $B_{k,\mathrm{U}}$, for both targets and UEs, respectively. The formulated $\mathcal{P}_{1}$ is a compact optimisation problem for our system model, while the non-compact problem can be found in \eqref{eq:op2}, denoted as $\mathcal{P}_{2}$. This optimisation problem maximises $\mathrm{KLD}_{\mathrm{avg}}$ by optimally allocating the power and the number of antennas among UEs and targets in both systems while maintaining a certain minimum KLD for each. Because the antennas are allocated through the precoding process, they can be allocated per subsystem, whereas a certain amount of power can be allocated to each target and UE.

\begin{table*}
\begin{minipage}{1\textwidth}
\begin{equation}
\begin{aligned}
%\mathcal{P}_{2}:\text{ }\text{ }\text{ }\text{ }\text{ }\text{ }\text{ }\text{ }\text{ }\text{ }&&&\\
\mathcal{P}_{2}: \max_{P_{\mathrm{c},k},P_{\mathrm{r},t},N_{\mathrm{c}},N_{\mathrm{r}}} \quad &\!\!\!\! \left(\mathrm{c_{com}}\:\sum_{k=1}^{K}{\frac{M \:(N_\mathrm{c}-K+1)\:P_{\mathrm{c},k}\:d_{\mathrm{c},k}^{-\eta}}{2\left( M-1\right) \left(P_{\mathrm{r}}\: \sigma _{F}^{2}\:d_{\mathrm{c},k}^{-\eta}+\sigma _{n}^{2}\right)\ln2}}\!\right.+\!\!\:\mathrm{c_{rad}}\sum_{t=1}^{T}{\!\!\frac{1}{4}\!\left(\! 1.4427\lambda _{t}\!-\!\frac{1}{\ln 2}
\!\!\int\limits_{0}^{\infty }\!\!\!\mathrm{e}^{-0.5\xi }\ln \left( I_{0}\left(\!\! \sqrt{
\lambda _{t}\xi }\!\right) \!\right)\! \right)\!\! d\xi} \\
&\!+\left.\frac{1}{2}\left(\frac{-0.5\lambda _{t}}{\ln 2}+\frac{\mathrm{e}^{-0.5\lambda _{t}}}{2\ln 2}\int\limits_{0}^{\infty }\mathrm{e}^{-0.5\xi }I_{0}\left( \sqrt{\lambda _{t}\xi }\right) \ln \left( I_{0}\left( \sqrt{\lambda _{t}\xi }\right) \right) d\xi\right)\right)\\
\textrm{s.t.} \quad\quad\quad\quad & \text{\eqref{10.b}, \eqref{10.c}, \eqref{10.d}, and \eqref{10.e}}
  \label{eq:op2}\\
  \\
\mathcal{P}_{4}:\text{ }\text{ }\text{ }\text{ }\text{ }\text{ }\text{ } \max_{P_{\mathrm{c},k},P_{\mathrm{r},t}} \quad &\!\!\!\! \:\Bigg(\mathrm{c_{com}}\sum_{k=1}^{K}{\frac{\lambda \:\alpha _{\mathrm{ZF}}^{2}\:P_{\mathrm{c},k}\:d_{\mathrm{c},k}^{-\eta}}{2\:M\left( M-1\right) \left(P_{\mathrm{r}}\: \sigma _{H}^{2}\:d_{\mathrm{c},k}^{-\eta}+P_{\mathrm{r}}\: \sigma _{F}^{2}\:d_{\mathrm{c},k}^{-\eta}+\sigma _{n}^{2}\right)\ln2}}\\
&+\mathrm{c_{rad}}\sum_{t=1}^{T}{\frac{|\alpha _{t}|^2\:P_{\mathrm{r},t}\:d_{\mathrm{r},t}^{-\eta}\:|a(\theta_t)\:\mathrm{R_x}\:a(\theta_t)^H|^2}{2\:ln(2)\:(\sigma_{\mathrm{n}}^2+\sigma_{\mathrm{err_1}}^2\:d_{\mathrm{r},t}^{-\eta}\:\left(\sigma_{\mathrm{w}}^2\:N\:\mathrm{P}_\mathrm{c}+\:\mathrm{P}_\mathrm{r}\right)+\sigma_{\mathrm{err_2}}^2\:d_{\mathrm{r},t}^{-\eta}\:\sigma_{\mathrm{w}}^2\:N\:\mathrm{P}_\mathrm{c})}}\Bigg)\\
\textrm{s.t.} \quad\quad\quad\quad & \text{\eqref{10.b}, \eqref{10.c}, and \eqref{10.d}} 
\end{aligned}
\end{equation}
\vspace{-0.15in}
\medskip
\hrule
\vspace{-0.24in}
\end{minipage}
\end{table*}

The discrete nature of antenna allocation and the continuous nature of power allocation make optimisation challenging. This problem can be formulated as a constrained MINLP problem, which is known to be computationally expensive to solve \cite{Kumar2020MethodsAS}. One possible approach for solving this problem is using the GA approach, a metaheuristic optimisation technique inspired by natural selection and genetics. GA generates a population of candidate solutions and iteratively improves them by applying operations such as mutation, crossover, and selection. This process continues until a stopping criterion is met, such as reaching a maximum number of iterations or achieving a satisfactory solution. GA can provide a global optimum, but it requires significant computational power resources and may not be suitable for real-time applications \cite{10.5555/534133}.

Therefore, in this paper, we propose RIPM which is a low-complex heuristic and very effective algorithm. The proposed RIPM manages to provide comparative results with much less computational complexity when compared to GA. RIPM algorithm aims at solving the antenna and power allocation problem by exploiting an IPM with integer constraints as a base structure. Accordingly, RIPM treats the discrete nature of antenna allocation as a continuous variable and uses a continuous optimisation approach to find the solution, where in each iteration, the number of antennas $N_\mathrm{r}$ is rounded and the number of communication antennas is found as $N_\mathrm{c}=N-N_\mathrm{r}$. A pseudo-code for the introduced RIPM is provided below in \textbf{Algorithm 1} which explains the computations executed to realise this algorithm. It is worth noting that RIPM is an iterative algorithm that moves towards the solution by following a central path in the feasible region while satisfying the required constraints \cite{nocedal2006interior}. 

\begin{figure}[t]
\vspace{-0.1in}
\begin{algorithm}[H]
\caption{Rounding-based Interior-Point Method for MINLP}
\begin{algorithmic}[1]
\REQUIRE{$F(x)$: objective function defined in \eqref{10.a},\\ 
$g(x)$: linear and nonlinear constraints defined in \eqref{10.b},\eqref{10.c},\eqref{10.d}, and \eqref{10.e},\\
$x=\{P_{\mathrm{c},k},P_{\mathrm{r}ad,t},N_{\mathrm{c}},N_{\mathrm{r}}\}$: continuous and integer optimisation variables}
\STATE{Set $\epsilon_{\mathrm{opt}} = 10^{-6}$, $\epsilon_{\mathrm{constr}} = 10^{-6}$, and $\mathrm{max\_iter} = 1000$}

\STATE{Define a modified objective function $F_m(x)$:}
\STATE{~~~Function $F_m(x)$:}
\STATE{~~~~~~Round integer variables in $x$ to their nearest integer \text{ }\text{ }\text{ }\text{ }\text{ } values, obtaining $x_r$}
\STATE{~~~~~~Compute $F(x_r)$ using the original objective function \text{ }\text{ }\text{ }\text{ }\text{ } $F(x)$}
\STATE{~~~~~~Return $F(x_r)$}
\STATE{~~~End Function}
\STATE{Relax integer variables in the MINLP problem, creating a relaxed problem}
\STATE{Initialize IPM parameters and starting point $x^0$ for the relaxed problem}
\STATE{Replace the original objective function $F(x)$ with the modified objective function $F_m(x)$ in the relaxed problem}
\vspace{-0.15in}
\STATE{Set iteration counter $k \gets 0$}
\REPEAT
  \STATE{Apply the IPM to the relaxed problem with the modified objective function, updating the solution $x^k$}
  \STATE{Compute $||x^{k} - x^{k-1}||$, the optimality tolerance $\mathrm{opt\_tol}$ and constraint tolerance $\mathrm{constr\_tol}$ for the current solution $x^k$}
  \STATE{$k \gets k + 1$}
\UNTIL{$(\mathrm{opt\_tol} < \epsilon_{\mathrm{opt}}$ and $\mathrm{constr\_tol} < \epsilon_{\mathrm{constr}})$ or $k \geq \mathrm{max\_iter}$}
\STATE{Obtain the final solution $x^*$ of the relaxed problem with the modified objective function}
\STATE{Round the integer variables in $x^*$}
\RETURN{$x^*$}
\end{algorithmic}
\end{algorithm}
\vspace{-0.5in}
\end{figure}

\subsection{Shared Deployment}
In shared deployment, only the power is optimally allocated as both systems utilizes the whole antenna array, the optimisation problem can be formulated similar to $\mathcal{P}_{1}$, as follows,

\begin{subequations}
\begin{align}
\mathcal{P}_{3}:\max_{P_{\mathrm{c},k},P_{\mathrm{r},t}} \quad\!\! & \mathrm{KLD}_{\mathrm{avg}}(P_{\mathrm{c},k},P_{\mathrm{r},t}) \label{11.a}\\
\textrm{s.t.}\quad &  A_{t,\mathrm{L}}\leq{\mathrm{{KLD}}_{\mathrm{r},t}}\leq A_{t,\mathrm{U}},  \,\,\,  \forall t \in T,  \label{11.b}\\
  &B_{k,\mathrm{L}}\leq{\mathrm{{KLD}}_{\mathrm{c},k}}\leq B_{k,\mathrm{U}},      \forall k \in K,   \label{11.c}  \\
  & \sum_{k=1}^{K}{P_{\mathrm{c},k}}+\sum_{t=1}^{T}{P_{\mathrm{r},t}}= P_{\mathrm{T}}.  \label{11.d}
\end{align}
\end{subequations}

The non-compact problem can be found in \eqref{eq:op2}, denoted as $\mathcal{P}_{4}$. This problem is solved using the IPM algorithm, which has a very low computational complexity. It is important to note that the shared deployment optimisation problem has less complexity from the separated deployment optimisation problem and that is due to the fact that only power allocation is required for the shared deployment, which makes this optimisation problem a NLP, rather than a MINLP. The results will show that not only the shared deployment has better performance but lower resource allocation complexity as well.

\section{Complexity analysis}

Optimization problems often necessitate the need for algorithms that provide a good trade-off between solution accuracy and computational efficiency. Complexity analysis gives a theoretical perspective on this, while empirical analysis provides practical insights. Here, we delve into the complexity of the RIPM and contrast it with that of GA.

\subsubsection{RIPM Complexity Analysis}

The RIPM introduces a hybrid approach, effectively capitalizing on the deterministic search strategy of IPM while incorporating a rounding technique adept for MINLP problems.

For $I$ iterations, RIPM's estimated complexity is:
\begin{equation}
     O(E + I \times (i + n^3 + (k+1) \times E + n + m)) ,
\end{equation}
where $E$ is the cost of evaluating the objective function and its derivatives, this is assumed to be of constant time complexity, $n$ is the total number of variables, $m$ is the number of constraints, $i$ is number of integer variables, $I$ is the number of IPM iterations and it is assumed with empirical verification that for the given problem domain, the RIPM and IPM will converge within a feasible number of iterations, and $k$ is the maximum number of objective function evaluations per iteration.

The complexity of RIPM predominantly escalates with the cube of the number of variables ($n^3$), contingent on dense matrix operations. Nevertheless, given the problem structure, matrix operations remain tractable, and the rounding technique, which could have been a potential source of discontinuities, operates seamlessly.

\subsubsection{GA Complexity Analysis}

GA are recognized for their heuristic nature and widespread search potential. The computational demands of GA, for $G$ generations with a population of size $P$, is estimated as \cite{10.5555/534133},
\begin{equation}
     O(G \times P \times L \times F(n)) ,
\end{equation}  
where $L$ is the length of the chromosome (often proportional to $n$), $F(n)$ is the time taken for a single fitness evaluation which is a function of the number of variables.

The complexity of GA is linearly related to the population size, and the generation count, also the $F(n)$ differs from problem to problem and its influence is more dominant in our problem. It's noteworthy that the probabilistic and explorative nature of GA, while powerful, can sometimes lead to broader, less-directed searches, which might not always be efficient.

When contrasting the complexities of RIPM and GA, it becomes apparent that the RIPM's structured and deterministic approach holds a tangible advantage. The RIPM complexity, while escalating polynomially with variables, remains moderated by the IPM's determinism and the efficient integration of the rounding technique. Meanwhile, the GA linear dependence on the population size and generations can render it demanding, especially for larger populations or extended generations. Our findings further accentuate this observation, showcasing RIPM's superior efficiency for the problem at hand.

\subsubsection{IPM}

The IPM for nonconvex NLP benefits from a systematic search strategy that offers determinism, even in the midst of nonconvexities. The IPM's complexity, for a given \( I \) iterations, can be depicted as \cite{nocedal2006interior},
\begin{equation}
     O(I \times (E + n^3 + m \times n^2)) ,
\end{equation}
where the variables have same definition for RIPM. Our empirical observations accentuate IPM's ability to approach solutions swiftly, minimizing the implied computational toll of the $n^3$ factor. GA has same complexity analysis for MINLP as in NLP but the single fitness evaluation will go higher in MINLP due to more variables and integers, GA heuristic fabric require exhaustive function evaluations, especially when precision is paramount. This inherent characteristic was underscored in our experiments, where GA fell out compared to IPM in execution rapidity and efficiency.

\section{Numerical Results}

This section presents the performance results of the above-introduced optimisation problem for both ISAC systems. The optimisation is carried out for single-UE-single-target (SUST) and MUMT scenarios. Mixed-integer GA and RIPM are compared for the first scenario only, e.g., SUST, as running the mixed-integer GA requires extremely high computational complexity for the case of MUMT, which makes it unsuitable for real-time applications compared to the RIPM. In the simulations, GA population size is set to $50$ times the number of decision variables, with a crossover fraction of $0.8$ and a mutation rate of $0.2$. The stopping criteria for GA include reaching the maximum number of generations, which is set to $100$ times the number of decision variables, or a stall in the best fitness value for a predetermined number of generations. While for the RIPM simulations, the stopping criteria for RIPM is based on the maximum number of iterations which is set to $1000$. Furthermore, the optimality tolerance and the constraint tolerance are set to $10^{-6}$.
For MUMT scenario, a number of $3$ UEs and $3$ targets have been considered for two cases; the first case without upper limits, whereas upper limits on individual KLDs have been enforced in the second one. In all cases, the antenna separation is set to half wavelength, the total transmit power is fixed at $P_\mathrm{T}=1$, the covariance matrix for the radar in the case of separated deployment is $\mathbf{R}_{x}=\mathbf{I}_{N_\mathrm{r}}$, and in the case of shared deployment $\mathbf{R}_{x}=\mathbf{I}_{N}$, the target cross section is normalised to $\alpha_t=1\:\forall t$, and QPSK modulation is used throughout the results. Moreover, the variance of the channel estimation error of the IC process is $\sigma_{\mathrm{err}}=\sigma_{\mathrm{err_1}}=0.01$, the pathloss exponent is $\eta=3$, which is considered to model the effect of large-scale fading, and the total number of antennas at BS is fixed at $N=20$. In the figures provided, the $\mathrm{P}_\mathrm{R}/{N}_o$ is chosen, so the optimisation problem is in the feasible region. Unless stated otherwise, $\mathrm{c_{com}}=\mathrm{c_{rad}}=\frac{1}{K+T}$.
\vspace{-0.01in}
\subsection{Separated Deployment}
\begin{figure}[ht]
\centering
\vspace{-0.1in}
\includegraphics[width=3.3in]{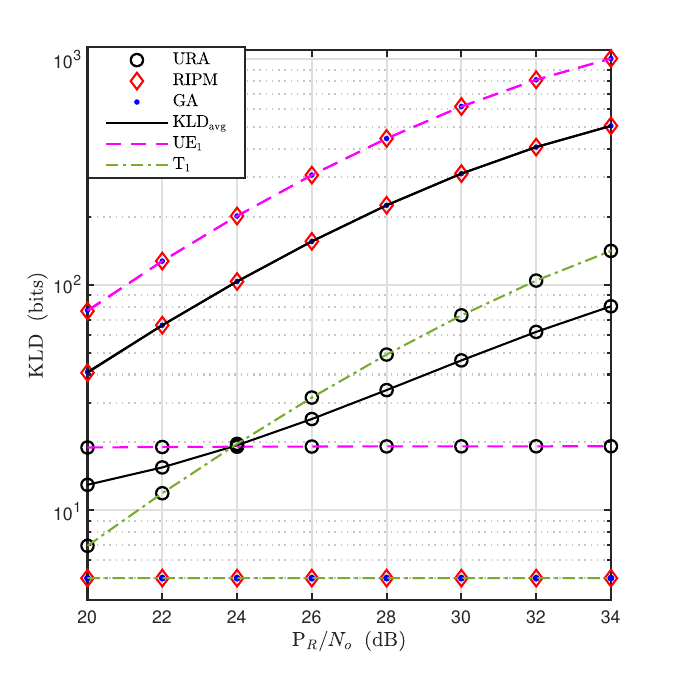} 
\vspace{-0.15in}
\caption{RIPM and GA optimisation methods comparison for SUST scenario, the minimum KLD requirement for the UE and target are $B_{1,\mathrm{L}}= 20$ and $A_{1,\mathrm{L}}=5$, respectively. The distances are $d_{\mathrm{UE}_1}=150$ m, and $d_{t_1}=220$ m. }
\label{fig:1}
\vspace{-0.01in}
\end{figure}

\begin{figure*}[!h]
\centering
\vspace{-0.3in}
\hspace*{-1cm} 
\includegraphics[width=7.5 in]{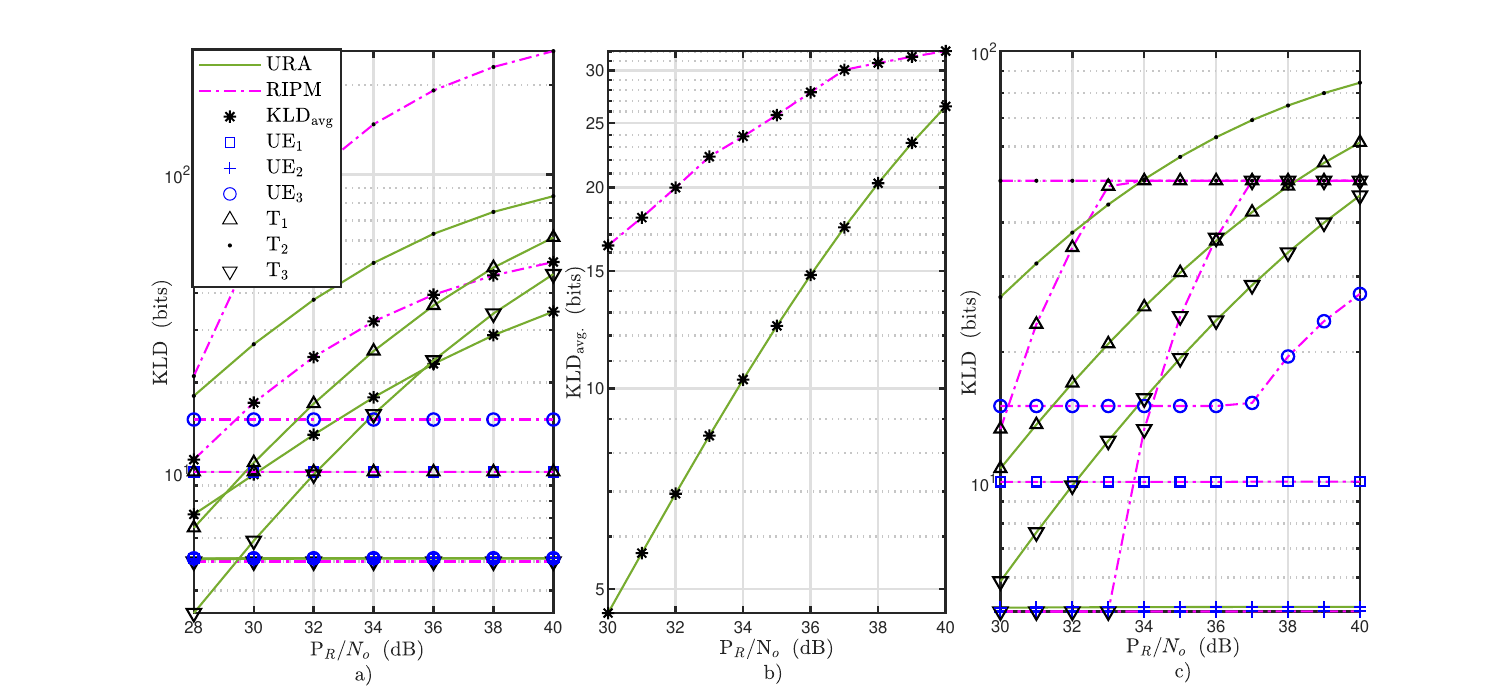}
\vspace{-0.2in}
\caption{RIPM optimisation is used for MUMT scenario, the minimum KLD for each UE is $B_{1,\mathrm{L}}= 10$, $B_{2,\mathrm{L}}=5$, and $B_{3,\mathrm{L}}=15$, and for each target is $A_{1,\mathrm{L}}=10$, $A_{2,\mathrm{L}}=15$, and $A_{3,\mathrm{L}}=5$, where $d_{\mathrm{UE}_{1}}=150$ m, $d_{\mathrm{UE}_2}=210$ m, $d_{\mathrm{UE}_3}=100$ m, $d_{t_1}=250$ m, $d_{t_2}=180$ m, and $d_{t_3}=300$ m. }
\label{fig:2}
\vspace{-0.1in}
\end{figure*}

In Fig. \ref{fig:1}, $\mathrm{KLD}_{\mathrm{avg}}$ and the individual KLD for both the UE and the target are plotted using both RIPM and GA methods for optimisation. As can be seen from the figure, $\mathrm{KLD}_{\mathrm{avg}}$ for both optimisation methods is much higher than the $\mathrm{KLD}_{\mathrm{avg}}$ of URA. Therefore, both optimisation methods show significant improvement over URA method by maximising the $\mathrm{KLD}_{\mathrm{avg}}$ and satisfying the minimum KLD requirements. In addition, RIPM with integer constraint rounding and GA performances are identical. It is important to note that because the UE is closer to BS and has a higher minimum KLD requirement than the radar target, it is allocated the rest of the resources after the radar target's minimum KLD requirement is satisfied.

As shown in Fig. \ref{fig:2}.a, the system is generalised to multiple UEs and targets using the RIPM optimisation method. The performance of GA has not been considered in this figure because it is computationally untraceable for this scenario. It is evident from the figure that $\mathrm{UE}_1$ and $\mathrm{UE}_3$ for the URA case are in an outage for the whole range of $\mathrm{P}_\mathrm{R}/N_o$ as they do not satisfy the minimum KLD requirement, as well as, targets $\mathrm{T}_1$ and $\mathrm{T}_3$ are in an outage at $\mathrm{P}_\mathrm{R}/N_o = 28$ dB. The optimal $\mathrm{KLD}_{\mathrm{avg}}$ improvement over the URA case is still significant for MUMT scenario. It is important to note that changing the KLD requirements will change the gained improvement, as the minimum KLD requirement for UEs and targets affects $\mathrm{KLD}_{\mathrm{avg}}$. It will decrease whenever a higher minimum requirement is enforced on distant targets or UEs. In addition, Fig. \ref{fig:2}.a shows that the minimum KLD requirement is satisfied for all UEs and targets, and the rest of the resources are given to the target or UE that will provide the highest $\mathrm{KLD}_{\mathrm{avg}}$. In the considered scenario, $\mathrm{T}_2$ is the target that provides the highest $\mathrm{KLD}_{\mathrm{avg}}$, which is the closest target to BS, even though $\mathrm{UE}_3$ and $\mathrm{UE}_1$ are closer to BS than $\mathrm{T}_2$, this is because the communication system is very sensitive to radar-interference as there is no IC applied at UEs.

\begin{comment}

\begin{figure}[!h]
\centering
\vspace{-0.12in}
\includegraphics[width=3.3in]{F3_V1.pdf}
\caption{RIPM optimisation is used with an upper KLD limit equals $50$}
\label{fig:3}
\vspace{-0.1in}
\end{figure}
\end{comment}

In Fig. \ref{fig:2}.b and Fig. \ref{fig:2}.c, the same MUMT setup as Fig. \ref{fig:2}.a is used with the addition of upper limits applied on the KLD of each of UEs and targets. The upper limits are introduced to ensure fairness between UEs and targets. Once $\mathrm{T}_2$, the device with the highest KLD, as shown in Fig. \ref{fig:2}.a, reaches its KLD upper limit, which is set to $50$ bits for both $B_{k,\mathrm{U}}$ and $A_{t,\mathrm{U}}$ in this simulation setup, the extra resources will be allocated to the other UEs and targets in the order of their contribution in the maximisation of $\mathrm{KLD}_{\mathrm{avg}}$. The KLD for $\mathrm{UE_1}$ and $\mathrm{UE_3}$ in the case of URA are in outages as they are below the minimum KLD requirement.

\begin{comment}
In Fig. \ref{fig:4}, the receiver operating characteristic (ROC) curve is shown for all the targets, it is clear from the figure that the performance of the OA is superior to the URA, and it also shows that, as seen in the KLD curves $\mathrm{T_2}$ has the best probability of detection trade-off with the probability of false alarm due to it being the closest and being appointed more resources. This reinforces the usage of the KLD as a unifying metric for the calculations. Furthermore, the OA BER for each UE at $\mathrm{P}_\mathrm{T}/N_o=40$ dB are  $0.0175, 0.0620,$ and $0.0009$, respectively, and for URA BER for each UE are all the same at $0.0610$, this also shows how the BER performance for OA is more superior than the URA, as the latter suffers from an error floor.
\end{comment}

\subsection{Shared Deployment}

\begin{figure*}[ht]
\centering
\vspace{-0.2in}
\hspace*{-1cm} 
\includegraphics[width=7.4in]{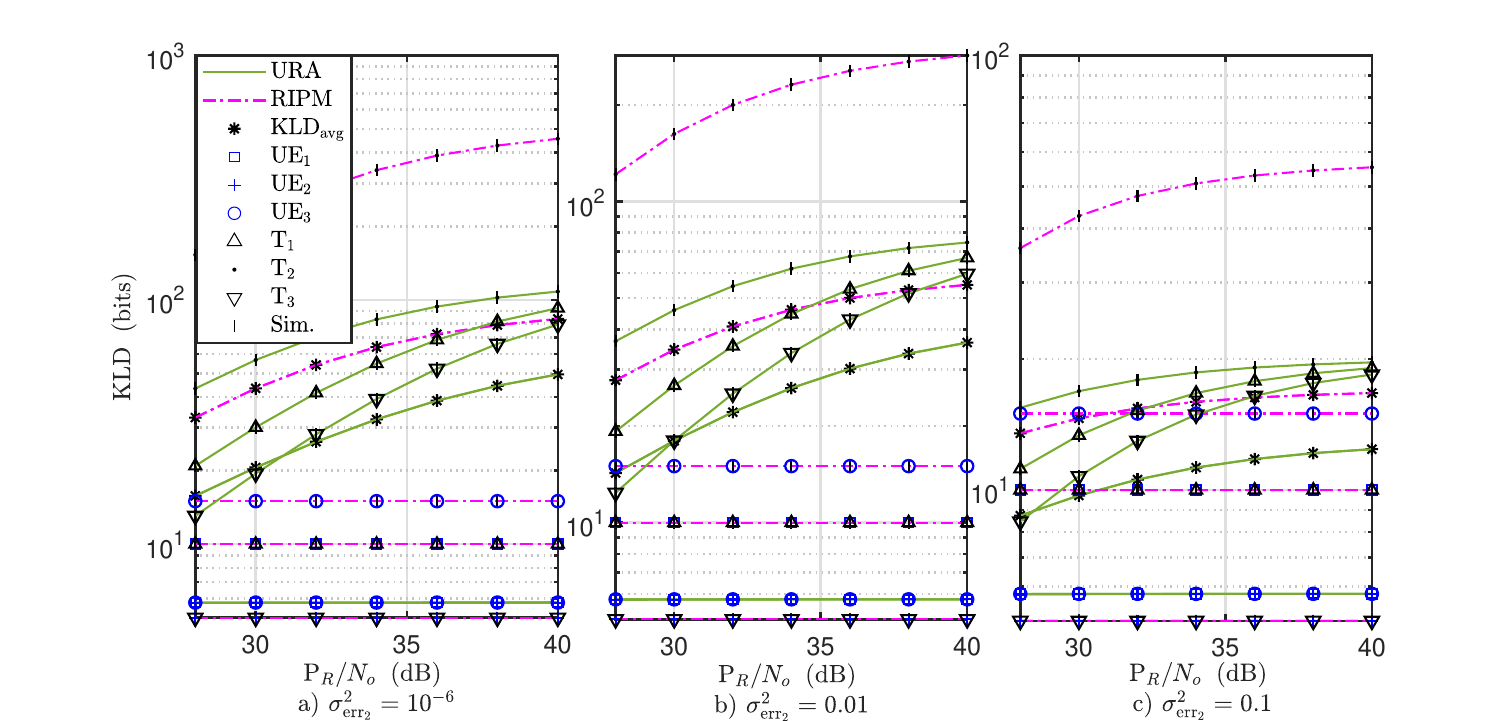} 
\vspace{-0.13in}
\caption{RIPM optimisation is used for shared deployment MUMT scenario, the configuration is the same as Fig.3, it is done for multiple cases of $\sigma_{\mathrm{err_2}}^2$ }
\label{fig:shared}
\vspace{-0.1in}
\end{figure*}

The shared deployment MUMT ISAC system power is optimally allocated by utilising the KLD in (19). Fig.\ref{fig:shared} shows that the performance of shared deployment is superior to the performance of separated deployment when the $\sigma_{\mathrm{err_2}}^2$ is in reasonable range, at the point $\mathrm{P}_\mathrm{R}/N_o=30$ dB, the separated deployment $\mathrm{KLD_{avg.}}$ is approximately 30 bits, while in the case of $\sigma_{\mathrm{err_2}}^2=0.01$, the $\mathrm{KLD_{avg.}}$ is approximately 45 bits. This shows that shared deployment system and moving to more integration of both systems is the way forward. The estimation error of radar parameter which is dictated by the error variance $\sigma_{\mathrm{err_2}}^2$, affects the amount of information received from the target which is reflected on radar KLD and consequently on the $\mathrm{KLD_{avg.}}$, more error leads to less information, and as shown in the figure this gives decrease in the KLD measure. It is also still evident from the figure that $\mathrm{UE}_1$ and $\mathrm{UE}_3$ for the URA case are in an outage for the whole range of $\mathrm{P}_\mathrm{R}/N_o$ as they do not satisfy the minimum KLD requirement, as the whole communication system in URA saturate at around 5.8 bits.

\begin{figure}[!h]
\vspace{-0.15in}
\centering
\includegraphics[width=3.2in]{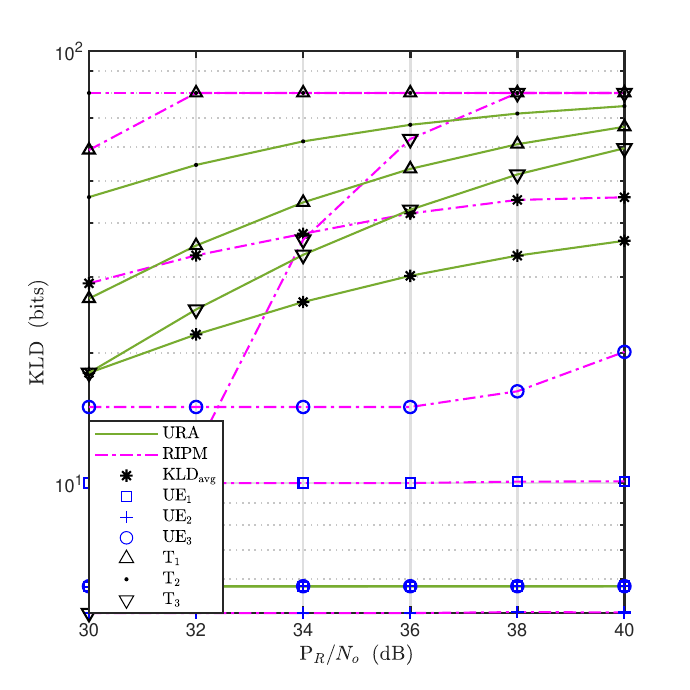}
\vspace{-0.1in}
\caption{RIPM optimisation is used for shared deployment MUMT scenario, the configuration is the same as Fig.3, with $\sigma_{\mathrm{err_2}}^2 = 0.01$ }
\label{fig:sh}
\vspace{-0.1in}
\end{figure}

In Fig. \ref{fig:sh}, the same setup as Fig. \ref{fig:shared}.b is used with the addition of upper limits applied on the KLD of each of UEs and targets. The upper limits are introduced to ensure fairness between UEs and targets just as shown in \ref{fig:2}.b and Fig. \ref{fig:2}.c for the separated deployment. This is done for shared deployment and the same thing is noticed where once $\mathrm{T}_2$, the device with the highest KLD, reaches its KLD upper limit, which is set to $80$ bits for both $B_{k,\mathrm{U}}$ and $A_{t,\mathrm{U}}$ in this simulation setup, the extra resources will be allocated to the other UEs and targets in the order of their contribution in the maximisation of $\mathrm{KLD}_{\mathrm{avg}}$.

\begin{figure}[!h]
\vspace{-0.35in}
\centering
\includegraphics[width=3.2in]{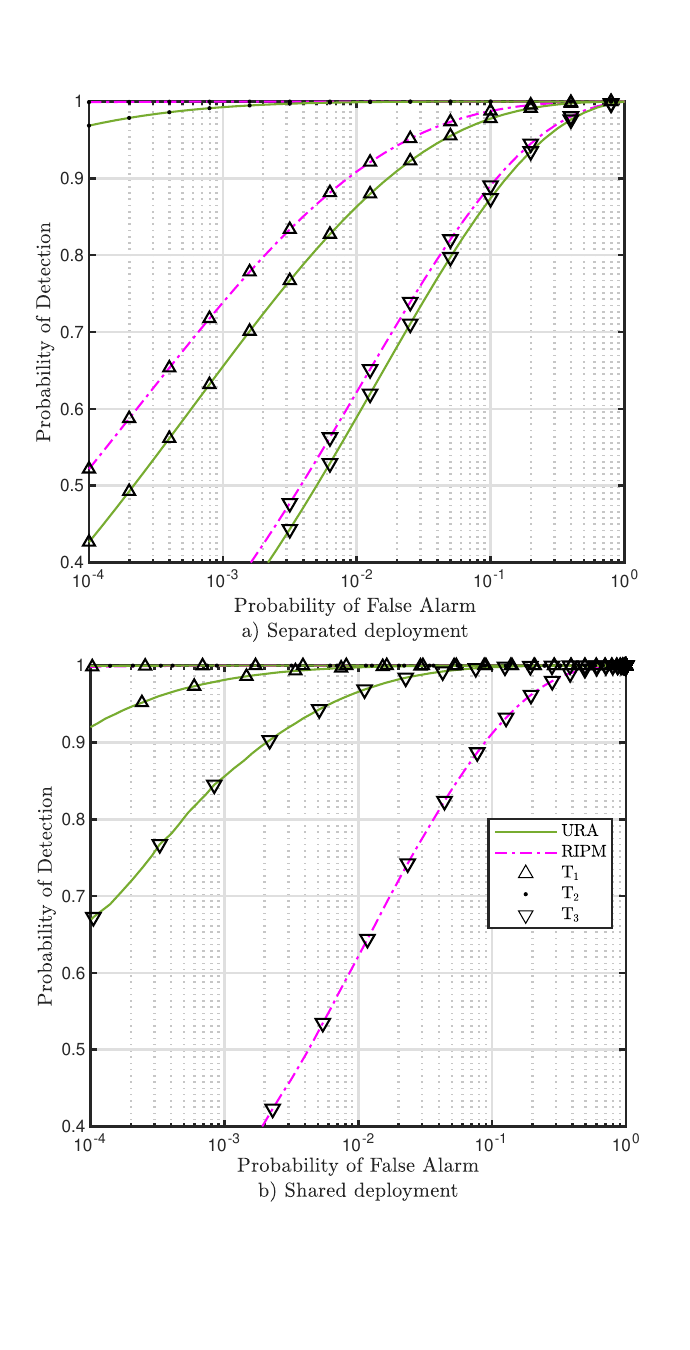}
\vspace{-0.6in}
\caption{ROC curves for the MUMT scenario with upper limit of 80 bits using RIPM and URA, at $\mathrm{P}_\mathrm{R}/N_o = 28$ dB. (a) Separated deployment; (b) Shared deployment.}
\label{fig:4}
\vspace{-0.2in}
\end{figure}

\vspace{0.1in}
For the sake of completeness, the receiver operating characteristic (ROC) curves are shown in Fig. \ref{fig:4} for all the targets using the same MUMT parameters in Fig.\ref{fig:2}.a but with an upper limit of 80 bits, this is done for both separated and shared deployment antenna configuration. It is clear from the figure that the performance of RIPM is superior to URA in terms of radar detection capability, for all targets in the separated deployment and for targets $\mathrm{T_1}$ and $\mathrm{T_2}$  for the shared deployment, this is because the URA basically only focuses on one system which is the one least susceptible to interference while the other system is totally cut-off, this is why a power optimisation is essential to ensure that both systems work properly, the upper limit introduced makes sure more fairness to not single out one UE or target.

As seen from the figure, $\mathrm{T_2}$ has the best probability of detection trade-off with the probability of false alarm because it is the closest target and allocated more resources. Furthermore, the bit error rates (BERs) of UEs achieved by using RIPM for both deployments at $\mathrm{P}_\mathrm{R}/N_o=40$ dB for both deployment configurations are  $0.0175, 0.0620,$ and $0.0009$, respectively, whereas a BER of $0.0610$ for all UEs is obtained using URA, it is important to note that the BER for each UE is determined by the set KLD minimum requirement for that UE, which can be changed to accommodate different BER requirements. By comparing the results in this figure with those in Fig.\ref{fig:2}.a, it can be realised that KLD measure can effectively characterise ISAC system. This sort of characterisation can be observed by the fact that a higher KLD results in better detection capability for the radar subsystem and lower BER for the communication subsystem. Apparently, the obtained results in this figure reinforce the fact that the usage of KLD as a unifying and homogeneous measure for ISAC system that provides a clear system level trade-off between both subsystems, and thus it can be effectively utilized as a standard measure for ISAC system design purposes.

\begin{figure}[!h]
\vspace{-0.35in}
\centering
\includegraphics[width=3.2in]{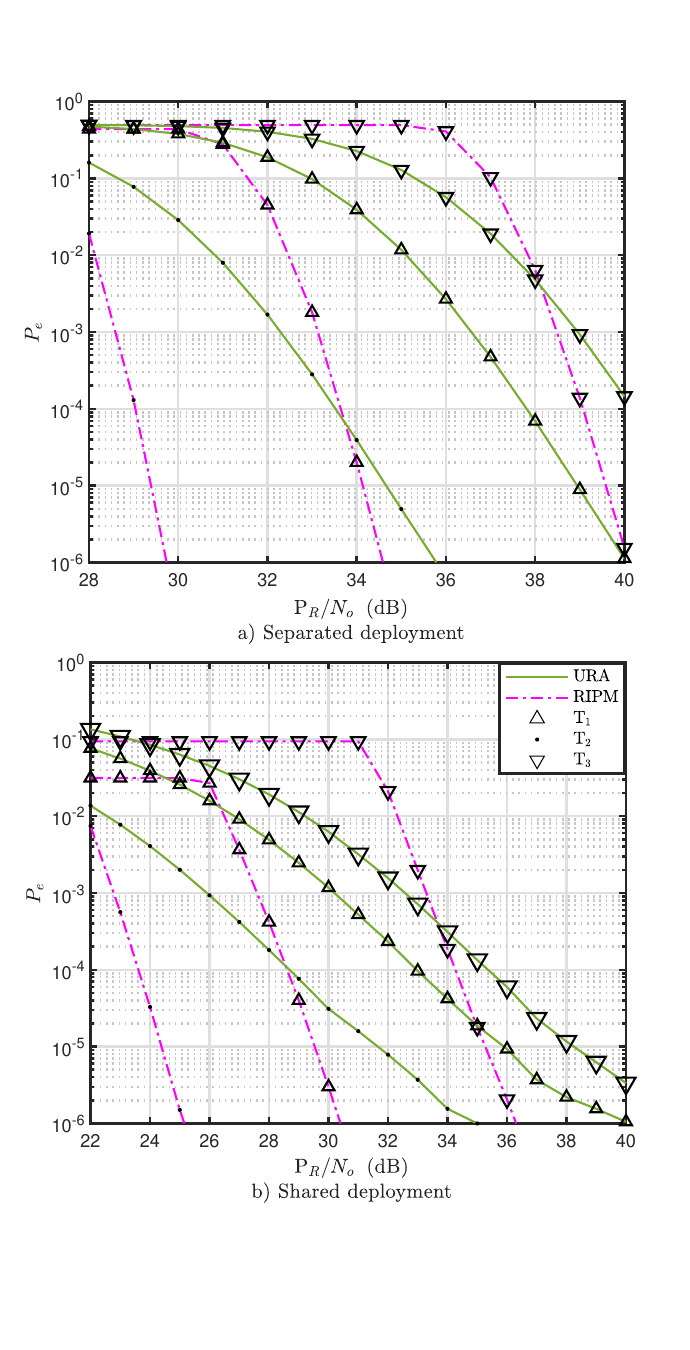}
\vspace{-0.6in}
\caption{Probability of error for the radar in the MUMT scenario with upper limit of 80 bits using RIPM and URA. (a) Separated deployment; (b) Shared deployment.}
\label{fig:error}
\vspace{-0.2in}
\end{figure}

In Fig. \ref{fig:error}, the probability of error can be observed for the radar in both antenna deployment configurations, the optimisation works to increase the total $\mathrm{KLD_{avg.}}$, satisfying the constraints, minimum KLD requirements, and the upper KLD fairness constraint. $T_2$ is the nearest between the targets and UEs, therefore, it will contribute the highest to the $\mathrm{KLD_{avg.}}$, but to ensure it is not singled out, the upper limit is introduced. In Fig. \ref{fig:error}.b, the upper limit is reached for $T_2$ at 26 dB, then the extra power resources will go to the next target $T_1$, and when that target reaches the upper limit the extra resources will go to the next target $T_3$ at around 31 dB, similar thing happens in Fig.\ref{fig:error}.a. but the upper limit for each target is reached at a higher $\mathrm{P}_\mathrm{R}/N_o$ which shows that the shared deployment configuration has superiority over the separated deployment. The URA can be a bit better than the RIPM for $T_1$ and $T_3$ before the point that they are given more resources, because the URA does not have any upper or lower limit for either the radar or communication systems, and the optimisation focuses the minimum requirement for both systems and their equipments.

 \begin{figure}[!h]
\vspace{-0.1in}
\centering
\includegraphics[width=3.3in]{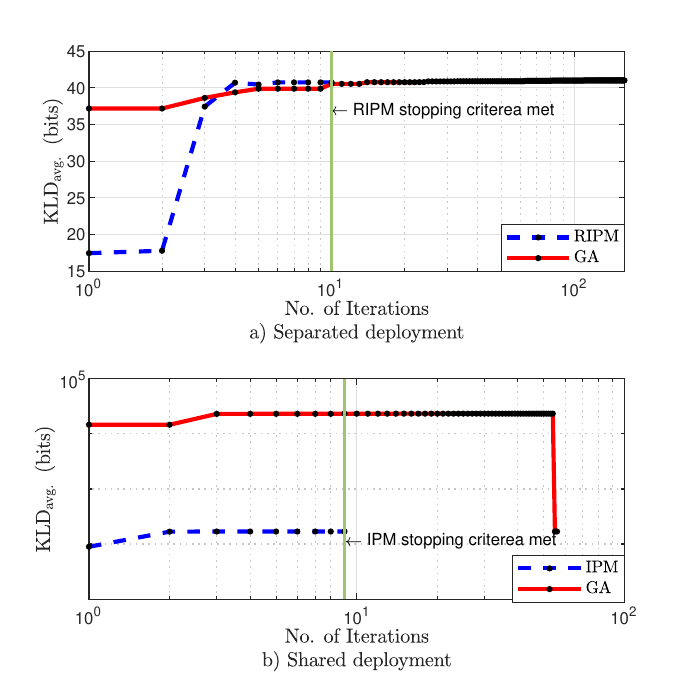}
\vspace{-0.2in}
\caption{Convergence evaluation for both RIPM and GA optimisation methods for both deployment configurations. (a) Separated deployment for SUST; (b) Shared deployment for SUST.}
\label{fig:conv1}
\vspace{-0.02in}
\end{figure}

For the sake of computational complexity argument, the convergence, optimality, and complexity of both optimisation methods, i.e., GA and RIPM, are investigated, as well. The time and computational resources consumed by GA are generally much more than RIPM due to the nature of its search process (e.g. selection, crossover, mutation). In Fig. \ref{fig:conv1}, the convergence criteria is shown through plotting $\mathrm{KLD}_{\mathrm{avg}}$ versus the numbers of iterations that are required for the optimisation techniques to find the optimal solution. For Fig.\ref{fig:conv1}.a, the convergence is explored for separated deployment at $\mathrm{P}_\mathrm{R}/{N}_o = 20$ dB, using the same simulation environment used for SUST configuration in Fig. \ref{fig:1}, where the RIPM method requires $10$ iterations to find the solution, while GA needs $164$ iterations to find the same solution. This clearly depicts that there is a massive difference in terms of the number of iterations. Moreover, the number of times the objective function is evaluated during the optimisation process using GA is $6275$ calculations, whereas it is $51$ calculations for the case of RIPM. Obviously, RIPM is a significantly better candidate to achieve the allocation objective. Generally speaking, the quality of the final solution determines the optimality condition. It can be noticed from the figure that the final solution for GA is $41.007$ bits, whereas it is $40.736$ for RIPM, and thus the percentage difference between them is as little as $0.6631$\%. This shows that GA gives a very tiny improvement over the RIPM regarding the final solution. However, this tiny amount is not as proportional to the amount of the functions calculations and the iterations difference between the two methods.

In Fig. \ref{fig:conv1}.b, the shared deployment convergence for both the GA and RIPM is shown, this is done at $\mathrm{P}_\mathrm{R}/{N}_o = 20$ dB, using the same simulation environment used for the SUST configuration in Fig. \ref{fig:1}. This shows that GA falls in terms of efficiency in the shared deployment as well, this is observed in the figure as the GA starts with infeasible points and only starts searching in the right direction after 100 iteration, reaching its feasible solution at around 140 iterations. The number of objective function calculations for GA reaches around 2520 calculations in total, while the RIPM reaches the solution at 9 iterations with only 27 objective function calculations.

\begin{figure}[!h]
\vspace{-0.17in}
\centering
\includegraphics[width=3.3in]{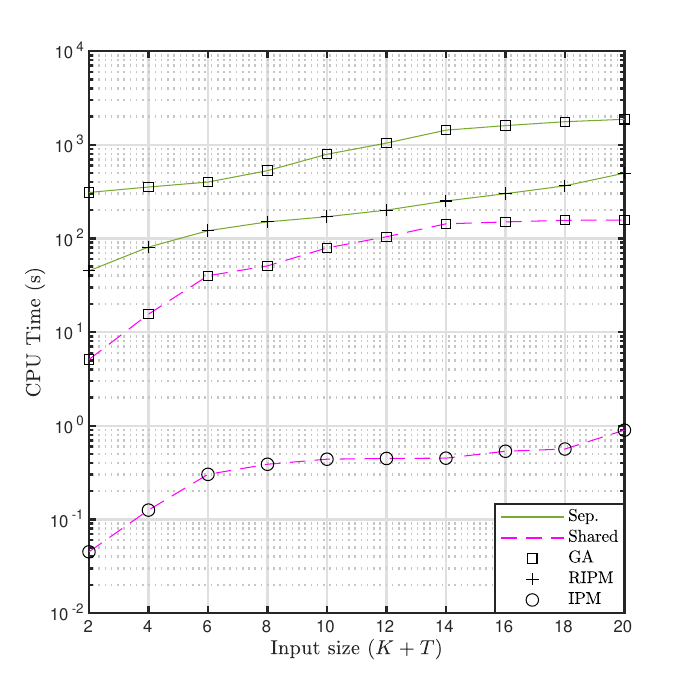}
\vspace{-0.2in}
\caption{CPU time simulation for both RIPM and GA optimisation methods for both deployment configurations.}
\label{fig:cpu}
\vspace{-0.05in}
\end{figure}

In Fig. \ref{fig:cpu}, the CPU time versus the input size which is $K+T$, this figure shows the average CPU time it takes to perform optimisation for both RIPM/IPM and GA in separated and shared deployment configurations by varying the input size i.e the number of continuous optimisation variables. This shows in both deployment configuration how the RIPM and IPM performs more efficiently in terms of time when compared with the GA technique, which also backs our convergence rate results.

\begin{figure}[!h]
\vspace{-0.17in}
\centering
\includegraphics[width=3.3in]{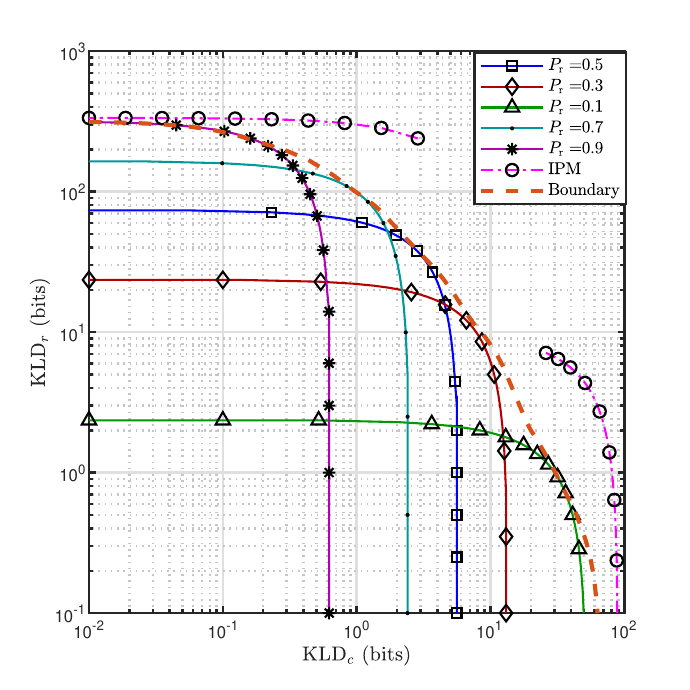}
\vspace{-0.1in}
\caption{ $\mathrm{KLD_{c}}$ versus $\mathrm{KLD_{r}}$ trade-off for shared deployment at $\mathrm{P}_\mathrm{R}/{N}_o = 44$ dB, same MUMT system configuration from Fig.\ref{fig:shared}.b, with relaxed constraints. }
\label{fig:to}
\vspace{-0.1in}
\end{figure}

Lastly, it is important to show the trade-off performance between the two systems in response to the weights of each system $\mathrm{c_{com}}$ and $\mathrm{c_{rad}}$ in \eqref{eq:13}. This kind of trade-off is achieved by varying $\mathrm{c_{com}}$ from 0 to $\frac{2}{T+K}$, calculating $\mathrm{c_{rad}}$ using $\mathrm{c_{rad}}=\frac{1-K\:\mathrm{c_{com}}}{T}$, and then plotting the achievable $\mathrm{KLD_{c}}$ against $\mathrm{KLD_{r}}$, where $\mathrm{KLD_{c}}=\mathrm{c_{com}}\sum_{k=1}^{K}{\mathrm{KLD}_{\mathrm{c},k}}$ and $\mathrm{KLD_{r}}=\mathrm{c_{rad}}\sum_{t=1}^{T}{\mathrm{KLD}_{\mathrm{r},t}}$. As can be observed from Fig. \ref{fig:to}, which depicts this trade-off for the shared deployment scenario, the IPM optimised method provides a better trade-off between the radar and communication system, and much more controllability when compared to the URA or any fixed power approach. When observing the boundary which is the maximum achievable performance trade-off using the fixed power approach and the IPM optimised trade-off, the optimisation clearly provides a much better trade-off in comparison. For example, the maximum achievable $\mathrm{KLD_{c}}$ in Fig. \ref{fig:to} is about $50$ bits achieved when $\mathrm{KLD_{r}}$ approaches 0, whereas a $\mathrm{KLD_{c}}$ of $90$ bits is obtained using IPM at $\mathrm{P}_\mathrm{R}/{N}_o = 44$ dB. Some sort of discontinuity can be seen in Fig.\ref{fig:to}, in the $x$-axis range of (3,21). This behaviour is attributed to the fact that optimisation prioritises a system over the other after satisfying the minimum KLD requirement to get the maximum KLD gain, this will make the system at a certain weight switching the priority from one system to the other to provide optimal KLD gain. This zone is considered infeasible and does not provide maximum performance gain. In Fig.\ref{fig:to}, it is shown that the IPM optimised method provides better trade-off between the radar and communication system.

\vspace{-0.1 in}
\section{Conclusion}
\begin{comment}

In this paper, the allocation of the available antennas and power at BS of an ISAC system was considered in downlink with the aim of optimising ISAC functionality. Unlike existing literature, this work adopted the KLD measure to unify both subsystems into one integrated system in order to allocate antennas and power resources. The proposed RIPM algorithm was compared with GA and URA approach, where MUMT scenario was investigated, in addition to a SUST case. The obtained results demonstrated that the proposed RIPM managed to provide comparable performance to the optimal GA with significant reduction in the computational complexity. Simulation results also revealed that both GA and RIPM are able to provide significant improvements over URA in terms of KLD, BER, and probability of detection. The system design with the unifying KLD measure, as shown in the results, provides an efficient, clearer and more cooperative design for ISAC systems.
\end{comment}
In this paper, we explored new means for the allocation of antenna and power resources in downlink ISAC system to optimize its functionality. Unlike existing literature, we applied the KLD metric as a unifying measure to merge both subsystems into a singular integrated system, thereby facilitating the allocation of antennas and power resources in a novel manner. It was shown that this unifying KLD measure can provide an efficient, clearer and more leaner ISAC design.

Both separated and shared ISAC modes of operation have been considered. The findings were decisive in showing that the shared deployment outperformed the separated one which suggests that deeper integration offers better rewards and hence a more promising direction for future ISAC systems.

Furthermore, the proposed RIPM algorithm was benchmarked against the GA and URA approaches. Notably, we delved into the MUMT scenario and a SUST one.  The obtained results demonstrated that the proposed RIPM provides comparable performance to the optimal GA with significant reduction in computational complexity. The results highlighted that both GA and RIPM have an edge over URA, especially in the realms of KLD, BER, and probability of detection.

\vspace{-0.05 in}
\bibliographystyle{IEEEtran}
\bibliography{Ref}

\end{document}